\documentclass[preprint,prd,
amsmath,amssymb,amsthm,nofootinbib,superscriptaddress,longbibliography]{revtex4-2}

\usepackage{graphicx}   
\usepackage[
colorlinks=true,        
citecolor=blue,         
linkcolor=blue,         
urlcolor=blue ,          
]{hyperref}  
\usepackage{tabulary}
\usepackage{color}      
\usepackage{orcidlink}
\usepackage{multirow}
\usepackage{floatrow}
\usepackage{float} 
\usepackage{amsmath}
\usepackage{bm}
\usepackage[stretch=10]{microtype}

\newcommand{\nc}{\newcommand*} 
\newcommand{\be}{\begin{equation}}
	\newcommand{\ee}{\end{equation}}
\newcommand{\bea}{\setlength\arraycolsep{2pt} \begin{eqnarray}}
	\newcommand{\eea}{\end{eqnarray}}

\nc{\al}{\alpha}
\nc{\s}{\sigma}
\nc{\dt}{\delta}
\nc{\Dt}{\Delta}
\nc{\Ld}{\Lambda}
\nc{\p}{\partial}
\nc{\om}{\omega}
\nc{\Om}{\Omega}
\nc{\rd}{\mathrm{d}}
\nc{\Od}[1]{\mathcal{O}(#1)} 
\nc{\kp}{\kappa}

\def\[{\left[}
\def\]{\right]}
\def\e{\begin{equation}}
	\def\q{\end{equation}}
\def\m{\begin{eqnarray}}
	\def\n{\end{eqnarray}}

\nc{\Eq}[1]{Eq.~\eqref{#1}}     
\nc{\Fig}[1]{Fig.~\ref{#1}}     
\nc{\Table}[1]{Table~\ref{#1}}  
\nc{\Sec}[1]{Sec.~\ref{#1}}     

\nc{\Msun}{M_\odot}             
\nc{\fpbhn}{f_{\mathrm{pbh0}}}    
\nc{\mR}{\mathcal{R}} 
\nc{\seq}{\sigma_{\mathrm{eq}}}
\nc{\ogw}{\Omega_{\mathrm{GW}}}
\nc{\gpcyr}{\mathrm{Gpc}^{-3}\,\mathrm{yr}^{-1}}
\nc{\lvc}{LIGO/Virgo} 
\nc{\SNR}{\mathrm{SNR}} 
\nc{\mmin}{{m_{\mathrm{min}}}}
\nc{\mmax}{{m_{\mathrm{max}}}}
\nc{\Mmin}{{M_{\mathrm{min}}}}
\nc{\fmin}{{f_{\mathrm{min}}}}
\nc{\VT}{\mathrm{VT}}
\nc{\rhoGW}{\rho_{\mathrm{GW}}}
\nc{\vth}{\vec{\theta}}
\nc{\vd}{\vec{d}}
\nc{\vla}{\vec{\lambda}}
\nc{\Nobs}{N_{\mathrm{obs}}}
\nc{\av}[1]{\langle #1 \rangle} 
\nc{\km}{\mathrm{km}}
\nc{\Mpc}{\mathrm{Mpc}}
\nc{\Tobs}{T_{\mathrm{obs}}}
\nc{\Ntemp}{N_{\mathrm{temp}}}
\nc{\ie}{\textit{i.e.}}
\nc{\eg}{\textit{e.g.~}}
\nc{\app}{\approx}
\nc{\hf}{\frac{1}{2}}

\newcommand{\apjl}{Astrophys. J. Lett.}
\newcommand{\apjs}{Astrophys. J. Suppl. Ser.}

\newcommand{\aap}{Astron. \& Astrophys.}
\newcommand{\aj}{Astron. J.}
\newcommand{\araa}{Ann. Rev. Astron. Astrophys.} 
\newcommand{\mnras}{Mon. Not. R. Astron. Soc.}
\newcommand{\jcap}{JCAP}

\newcommand{\physrep}{Physics Reports}

\newcommand{\aapr}{Astron. Astrophys. Rev.}

\def \pccm {\mathrm{pc\ cm^{-3}}}
\def \kmsMpc {\mathrm{km\ s^{-1}\ Mpc^{-1}}}

\begin{document}

\title{Constraining the Baryon Fraction in Extragalactic Diffuse Ionized Gas with 124 Localized Fast Radio Bursts}

\author{Yang Liu }
\affiliation{Department of Physics and Key Laboratory of Low-Dimensional Quantum Structures and Quantum Control of Ministry of Education, Hunan Normal University, Changsha, Hunan 410081, China}
\affiliation{Purple Mountain Observatory, Chinese Academy of Sciences, No. 10 Yuanhua Road, Nanjing 210023, China}

\author{Yuchen Zhang}
\affiliation{Department of Physics and Key Laboratory of Low-Dimensional Quantum Structures and Quantum Control of Ministry of Education, Hunan Normal University, Changsha, Hunan 410081, China}

\author{Jun-Jie Wei}
\email{jjwei@pmo.ac.cn}
\affiliation{Purple Mountain Observatory, Chinese Academy of Sciences, No. 10 Yuanhua Road, Nanjing 210023, China}
\affiliation{School of Astronomy and Space Sciences, University of Science and Technology of China, No. 96 JinZhai Road, Hefei 230026, China}

\author{Xue-Feng Wu}
\email{xfwu@pmo.ac.cn}
\affiliation{Purple Mountain Observatory, Chinese Academy of Sciences, No. 10 Yuanhua Road, Nanjing 210023, China}
\affiliation{School of Astronomy and Space Sciences, University of Science and Technology of China, No. 96 JinZhai Road, Hefei 230026, China}

\author{Hongwei Yu}
\email{hwyu@hunnu.edu.cn}
\affiliation{Department of Physics and Key Laboratory of Low-Dimensional Quantum Structures and Quantum Control of Ministry of Education, Hunan Normal University, Changsha, Hunan 410081, China}
\affiliation{Hunan Research Center of the Basic Discipline for Quantum Effects and Quantum Technologies, Hunan Normal University, Changsha, Hunan 410081, China}

\author{Puxun Wu}
\email{pxwu@hunnu.edu.cn}
\affiliation{Department of Physics and Key Laboratory of Low-Dimensional Quantum Structures and Quantum Control of Ministry of Education, Hunan Normal University, Changsha, Hunan 410081, China}
\affiliation{Hunan Research Center of the Basic Discipline for Quantum Effects and Quantum Technologies, Hunan Normal University, Changsha, Hunan 410081, China}


\begin{abstract}

Fast radio bursts (FRBs) are increasingly recognized  as powerful cosmological tools for constraining the baryon fraction in  extragalactic diffuse ionized gas, presenting a promising approach to address the missing baryon problem. In this paper, we  constrain  the baryon fraction in  extragalactic diffuse ionized gas ($f_\mathrm{d}$) utilizing the latest sample of 124 localized FRBs across three different cosmological models. Our analysis models the probability distribution of the extragalactic diffuse ionized gas dispersion measure with a form that accurately reproduces mock observations.  For a constant $f_\mathrm{d}$ model, we find that more than 90\% of baryons reside in the diffuse ionized gas phase.   This result is robust against the choice of dark-energy parametrization under the current combination of datasets, although the fitted cosmological parameters shift accordingly.  We also find that the inferred $f_\mathrm{d}$ is sensitive to the assumed dispersion measure distributions of both the Milky Way halo and the FRB host galaxies.  Furthermore,  the current data do not show statistically significant evidence for redshift evolution in $f_\mathrm{d}$, but the constraints are limited by the redshift distribution of the sample.  Our conclusions are insensitive to the adopted baryonic feedback parameters and to the dispersion measure selection effect. These results provide strong evidence that the majority of the missing baryons reside in the diffuse ionized intergalactic medium.
\end{abstract}
\maketitle

\section{Introduction}

Current cosmological measurements, such as the cosmic microwave background (CMB)~\citep{2020A&A...641A...6P} and  Big Bang nucleosynthesis (BBN)~\citep{2018ApJ...855..102C}, suggest that approximately 5\% of the Universe's energy budget  is composed of  baryonic matter. 
However, as first determined by \citet{1992MNRAS.258P..14P}, the total amount of visible baryons falls well below the value expected from standard primordial nucleosynthesis.
Current estimates show that the collapsed structures, including galaxies, groups, clusters, and the circumgalactic medium, account for only about 18\% of the expected baryon content~\citep{1998ApJ...503..518F,2004ApJ...616..643F,2012ApJ...759...23S}.  Observations of the photoionized Lyman-$\alpha$ (Ly$\alpha$) forest and the warm-hot intergalactic medium detect  about 53\% of  the baryons,  leaving roughly  30\% baryons unaccounted for~\citep{2012ApJ...759...23S}. This so-called ``missing baryon problem'' has persisted as a significant puzzle in cosmology for decades.
These missing baryons are believed to  be located in the diffuse intergalactic medium (IGM) and the galactic halos,  yet remain too faint for direct detection~\citep{2016ARA&A..54..313M}.

Fast radio bursts (FRBs), which are millisecond-duration radio transients, provide a promising tool for probing the missing baryons.   Their dispersion measures (DM) encode information about the ionized baryon density along the line of sight of the FRB signal~\citep{2007Sci...318..777L, 2013Sci...341...53T, 2019A&ARv..27....4P, 2019PhR...821....1P, 2023RvMP...95c5005Z, 2022A&ARv..30....2P}. 
For FRBs at cosmological distances, the observed DM incorporates a contribution from the extragalactic diffuse ionized gas ($\mathrm{DM_{cos}}$). This component  directly reflects   the ionized baryon density in the IGM, along with contributions from intervening halos. Consequently, FRBs have been extensively used  to probe the baryon fraction in the extragalactic diffuse ionized gas ($f_\mathrm{d}$) and to  constrain the present cosmic baryon density parameter $\Omega_{\mathrm{b0}}$~\citep{2014ApJ...783L..35D, 2019ApJ...872...88R, 2018PhRvD..98j3518M, 2019ApJ...876..146L, 2020MNRAS.496L..28L, 2020JCAP...04..023Q, 2019PhRvD.100j3519W, 2019JCAP...09..039W,2022ApJ...940L..29Y,2023ApJ...944...50W,2023MNRAS.520.6237L,2025arXiv250706841Z,2026arXiv260109521W}. By isolating the DM contribution from the baryonic component in the IGM and that from halos, FRBs can also be employed to investigate the baryon feedback processes in galaxies~\citep{2024MNRAS.529..537K, 2025NatAs...9.1226C}\:\footnote{ For 
	other cosmological applications of FRBs, see~\cite{2014ApJ...788..189G, 2014PhRvD..89j7303Z, 2018ApJ...856...65W, 2018ApJ...860L...7W, 2020ApJ...901..130Z, 2020ApJ...903...83Z, 2022JCAP...02..006Q, 2016ApJ...830L..31Y, 2018NatCo...9.3833L, 2023ApJ...946L..49L, 2022MNRAS.511..662H, 2022MNRAS.516.4862J, 2025JCAP...01..018F, 2025FrASS..1273311P,2022arXiv221213433Z, 2023ApJ...955..101W, 2022MNRAS.515L...1W,  2025PDU....4801926K, 2024MNRAS.527.7861G, 2014ApJ...797...71Z, 2019MNRAS.485.2281C, 2021MNRAS.502.5134B, 2021MNRAS.502.2346H,	2015PhRvL.115z1101W, 2016ApJ...821L...2N, 2016ApJ...820L..31T, 2015PhRvL.115l1301M,2016ApJ...822L..15W, 2017PhRvD..95l3010S, 2025ApJ...981....9W, 2023MNRAS.520.1324L, 2023JCAP...09..025W, 2024MNRAS.533L..57K, 2016ApJ...824..105A, 2023MNRAS.526.1773F, 2025A&A...698A.215G, 2025JCAP...11..019L, 2024A&A...683A..71W, 2026ApJ...998..109S, 2025arXiv251009463J, 2023SCPMA..6620412Z, 2025SCPMA..6880406Z, 2026ApJ...998...15L, 2026arXiv260422105S, 2026RAA....26h4010W, 2026arXiv260629761Z}.}.
Recently,  Macquart et al. \cite{2020Natur.581..391M}  developed a maximum likelihood estimation~(MLE) method to infer the current cosmic baryon density parameter and demonstrated consistency with  CMB and BBN results,  using just five  localized FRBs  assuming a flat $\Lambda$CDM (cosmological constant dark energy + cold dark matter) cosmological model with fixed present matter density $\Omega_{\mathrm{m0}}$. Their findings suggested that FRBs could potentially resolve the missing baryon problem.  Later, Wang et al. \cite{2023ApJ...944...50W} and Yang et al. \cite{2022ApJ...940L..29Y}  applied this MLE method to larger samples of 17 and 22 localized FRBs, respectively, and obtained $f_\mathrm{d} = 0.927 \pm 0.075$ and $f_\mathrm{d} = 0.83 \pm 0.06$ in the $\Lambda$CDM model. 
More recently, Connor et al. \cite{2025NatAs...9.1226C} combined 39 newly localized FRBs from Deep Synoptic Array (DSA)-110 with 30 previously localized sources, resulting in a   constraint of $f_\mathrm{d} = 0.94^{+0.05}_{-0.05}$ within the $\Lambda$CDM framework.
Notably, $f_\mathrm{d}$ is expected to increase with redshift, as massive halos are less abundant in the early Universe~\citep{2014ApJ...780L..33M, 2019MNRAS.485..648P}. However, the analysis in~\citep{2023MNRAS.520.6237L} found no statistically significant evidence for such redshift evolution.  
It is important to highlight that the probability density function (PDF) of $\mathrm{DM_{cos}}$,  which has been widely used in previous studies to construct the FRB likelihood function, generally takes the form provided by~\cite{2020Natur.581..391M}. This form is characterized by a modified-normal distribution with a power-law tail. Several recent studies have indicated that the PDF from~\cite{2020Natur.581..391M} does not accurately reproduce the results from cosmological simulations~\citep{2025arXiv250707090K, 2025ApJ...989...81S, 2024A&A...683A..71W}.  Additionally, it has been  found that this formulation omits a crucial factor, resulting in a systematic bias exceeding $1\sigma$ in the inferred value of $\Omega_{\mathrm{b0}}$~\cite{2025PhRvD.112h3516Z, 2026ApJ...996...66Z}. Moreover,  the commonly used parameterization of $\sigma_{\mathrm{d}}$, which is related to the physical variance of $\mathrm{DM_{cos}}$ in the PDF in~\citep{2020Natur.581..391M}, may also be inaccurate~\citep{2025ApJ...989...81S, 2026ApJ...996...66Z}. To address these issues, \cite{2026PhRvD.113d3513Z} proposed a PDF of $\mathrm{DM_{cos}}$ that aligns more closely with cosmological simulations compared to the formulations  proposed in~\citep{2020Natur.581..391M,2025arXiv250707090K}.
These findings motivate us to reassess the baryon fraction in the extragalactic diffuse ionized gas using the PDF of $\mathrm{DM_{cos}}$ given in~\citep{2026PhRvD.113d3513Z}.

In our study, we analyze constraints on $f_\mathrm{d}$ using a catalog of 124 localized FRBs. Unlike most of previous works that fixed cosmological parameters, we allow all cosmological parameters, including $f_\mathrm{d}$, to vary freely. Our analysis integrates FRB data with other complementary cosmological observations: the PantheonPlus Type Ia supernova (SN~Ia) sample~\citep{2022ApJ...938..113S, 2022ApJ...938..110B}, the baryon acoustic oscillation (BAO) data from the Dark Energy Spectroscopic Instrument (DESI)~\citep{2025PhRvD.112h3515A}, and the Planck CMB measurements~\citep{2020A&A...641A...6P, 2020A&A...641A...5P, 2020A&A...641A...8P}. Given that the  DESI BAO data strongly favor dynamical dark energy~\citep{2025PhRvD.112h3515A}, we adopt the $w_0w_a$CDM model as our primary cosmological framework. For comparison, we also present results for the $\Lambda$CDM and $w$CDM models. Initially, we examine a constant $f_\mathrm{d}$ and then explore potential redshift evolution by introducing two parametric models for $f_\mathrm{d}(z)$.

This paper is organized as follows: In section~\ref{sec:2}, we will describe  the likelihood function of FRBs. Section~\ref{sec:3} outlines  other cosmological datasets, including SN~Ia, BAO, and CMB, used in our analysis. The results are presented in section~\ref{sec:4}, and our conclusions are summarized in section~\ref{sec:5}.

\section{FRB dispersion measures and likelihood function}\label{sec:2}

The observed DM of an extragalactic FRB, denoted as $\mathrm{DM_{obs}}$, consists of several additive contributions:
\begin{eqnarray}\label{eq:DMobs}
	\mathrm{DM_{obs}}=\mathrm{DM_{ISM}}+\mathrm{DM_{halo}}+\mathrm{DM_{cos}}(z)+\mathrm{DM_{host}}(z),
\end{eqnarray}
where the subscripts ``ISM'', ``halo'', ``cos'', and ``host'' denote the contributions to DM from the Milky Way interstellar medium (ISM), the Milky Way halo, the extragalactic diffuse ionized gas, and the FRB host-galaxy, respectively. To constrain $f_\mathrm{d}$ from FRB observations, we need to construct a likelihood function 
 that accounts for the statistical distributions of the various components in Eq.~(\ref{eq:DMobs}).

For the  halo contribution $\mathrm{DM_{halo}}$, we assume it follows a truncated Normal distribution:
\begin{eqnarray}
	P_{\mathrm{halo}}(\mathrm{DM_{halo}}) \propto 
	\left\{ 
	\begin{array}{ll}
		\frac{1}{\sqrt{2\pi}\sigma_\mathrm{halo}}\exp\left[ -\frac{\left( \mathrm{DM_{halo}} - \overline{\mathrm{DM}}_\mathrm{halo}  \right)^2}{2\sigma_\mathrm{halo}^2} \right] ,&\  \text{if}\ \mathrm{DM_{halo}^{min}}\leq \mathrm{DM_{halo}}\leq \mathrm{DM_{halo}^{max}}, \\
		0,&\ \text{else},
	\end{array}
	\right.
\end{eqnarray}
where $\mathrm{DM_{halo}^{min}}$ and $\mathrm{DM_{halo}^{max}}$ denote the lower and upper bounds, while $\overline{\mathrm{DM}}_\mathrm{halo}$ and $\sigma_\mathrm{halo}$ represent the mean and standard deviation, respectively.
Based on the constraints from the all-sky H{\footnotesize I} 21 cm emission map, O{\footnotesize VII} absorption, and the DM to the Large Magellanic Cloud with hydrostatic models of halo gas, $\mathrm{DM_{halo}}$ is  estimated  to be in the range of $50\text{--}80~\pccm$~\cite{2019MNRAS.485..648P}.
However, a recent study by \cite{2025AJ....169..330R}, based on FRB 20220319D, derived a conservative upper limit of $28.7$ or $47.3~\pccm$.
While these values are lower than   those reported by \cite{2019MNRAS.485..648P}, they represent only the upper limit along the specific line of  sight to FRB 20220319D.
Given the complexity and potential anisotropy of the distribution of ionized gas in the Galactic halo~\citep{2012ARA&A..50..491P,2017ARA&A..55..389T}, we consider two sets of parameters in $P_{\mathrm{halo}}$ in our analysis.
In the first case, we adopt $\overline{\mathrm{DM}}_\mathrm{halo}=65~\pccm$ and $\sigma_\mathrm{halo}=15$ within the range $[50, 80]~\pccm$ (hereinafter $\mathcal{N}(65,15^2)$), based on the findings of \cite{2019MNRAS.485..648P}.
In the second case, we assume $\overline{\mathrm{DM}}_\mathrm{halo}=35~\pccm$ and $\sigma_\mathrm{halo}=15$ within the range $[20, 50]~\pccm$ (hereinafter $\mathcal{N}(35,15^2)$),  based on the results of \cite{2025AJ....169..330R}.

For the Galactic ISM contribution ($\mathrm{DM_{ISM}}$), we also assume a truncated normal distribution over the range $\left[0, +\infty\right]$, with the mean value $\overline{\mathrm{DM}}_{\mathrm{ISM}}$ set by the estimate from the latest Galactic free electron density model, NE2025~\citep{2026arXiv260211838O}. The standard deviation is assumed to be $\sigma_{\mathrm{ISM}} = 0.3\,\overline{\mathrm{DM}}_{\mathrm{ISM}}$, motivated by the fact that  the NE2025 model shows a rms   fractional distance estimation uncertainty of  approximately  30\%.

The host-galaxy contribution $\mathrm{DM_{host}}$ is well described by a log-normal distribution~\citep{2020Natur.581..391M, 2020ApJ...900..170Z}:
\begin{eqnarray} \label{eq:P_host}
	P_{\mathrm{host}}(\mathrm{DM}_{\mathrm{host}})=\frac{1}{\sqrt{2\pi}\mathrm{DM}_{\mathrm{host}} s_{\mathrm{host}}}\exp\left[-\frac{(\ln\mathrm{DM}_{\mathrm{host}}-\mu_{\mathrm{host}})^{2}}{2 s_{\mathrm{host}}^{2}}\right].
\end{eqnarray}
Here parameters $\mu_\mathrm{host}$ and $s_{\mathrm{host}}$  depend on redshift and the morphological type of the FRB's host-galaxy.
Their values  can be estimated from the IllustrisTNG simulation. For repeating FRBs like FRB121102, repeating FRBs like FRB180916 and non-repeating FRBs, 
Zhang et al. obtained the best-fit values of parameters $\mu_\mathrm{host}$ and $s_{\mathrm{host}}$  at eight redshifts within the range of $z\in[0.1, 1.5]$~\citep{2020ApJ...900..170Z}. 
In our analysis, we classify host galaxies into three categories based on their FRB type and use cubic spline interpolation to estimate the values of $\mu_\mathrm{host}$ and $s_{\mathrm{host}}$ at the redshifts of individual FRBs. Additionally, we will also consider the scenario in which the rest-frame DM from the host contribution,  denoted as $\mathrm{DM}_{\mathrm{host,0}}=\mathrm{DM}_{\mathrm{host}}(1+z)$, follows a log-normal distribution with its distribution parameters  treated as free parameters.

The component $\mathrm{DM_{cos}}$ in Eq.~(\ref{eq:DMobs}) is strongly dependent on both  the cosmological model and the baryon density in the extragalactic diffuse ionized gas. Since the distribution of free electrons in the diffuse ionized gas  is inhomogeneous, the actual value of $\mathrm{DM_{cos}}$ fluctuates around its average value $\langle\mathrm{DM_{cos}}\rangle$.
The ratio of $\mathrm{DM_{cos}}$ and $\langle\mathrm{DM_{cos}}\rangle$ is commonly assumed to follow the distribution~\citep{2014ApJ...780L..33M, 2020Natur.581..391M,2021ApJ...906...49Z}:
\begin{eqnarray}\label{eq:P_cosmic}
	P_{\mathrm{\Delta}}(\Delta)=A\Delta^{-\beta}\exp\left[-\frac{(\Delta^{-\alpha}-C_0)^2}{2\alpha^2\sigma_{\mathrm{d}}^2}\right],\quad\Delta>0,
\end{eqnarray}
where $\Delta\equiv \mathrm{DM_{cos}}/\langle\mathrm{DM_{cos}}\rangle$, and the parameter $A$ is a normalization factor. 
In~\citep{2020Natur.581..391M}, both parameters $\alpha$ and $\beta$ were fixed to $3$, and $\sigma_{\mathrm{d}}$ was parameterized as $F z^{-0.5}$, where the parameter $F$ is related to the strength of the baryon feedback. 
However, recent work in~\cite{2025arXiv250707090K} demonstrated that a distribution of $\mathrm{DM_{cos}}$ similar to this form, with $\alpha=\beta=3$, provides a poor match to the IllustrisTNG simulations. The commonly used parameterization $\sigma_{\mathrm{d}} = F z^{-0.5}$ has also been found to be potentially inaccurate~\citep{2025ApJ...989...81S, 2026ApJ...996...66Z}. Since incorrect  $\sigma_{\mathrm{d}}$ form and parameter choices in Eq.~(\ref{eq:P_cosmic}) can bias the inferred value of $f_\mathrm{d}$, we do not impose any assumptions on the values of $\sigma_{\mathrm{d}}$, $\alpha$, and $\beta$. 
Instead, we fit these parameters in 53 redshift slices spanning $z = 0.02$ to $3$, using the simulated $\mathrm{DM_{cos}}$ data extracted from the IllustrisTNG simulations~\cite{2025arXiv250707090K}, following the procedure outlined in \citep{2026PhRvD.113d3513Z}. We then use cubic spline interpolation to obtain these parameter values at arbitrary FRB redshifts. The remaining parameter $C_0$ is fixed to $1$, and $A$ is determined as the normalization factor.
Importantly, Eq.~(\ref{eq:P_cosmic}) describes the distribution of the dimensionless variable $\Delta$, not directly that of $\mathrm{DM_{cos}}$. The corresponding PDF of $\mathrm{DM_{cos}}$ can be derived via a change of variables~\citep{2025PhRvD.112h3516Z}:
\begin{eqnarray} \label{eq:P_IGM}
P_{\mathrm{cos}}(\mathrm{DM_{cos}}) = \frac{1}{\langle\mathrm{DM_{cos}}\rangle} P_{\mathrm{\Delta}}\left(\frac{\mathrm{DM_{cos}}}{\langle\mathrm{DM_{cos}}\rangle}\right).
\end{eqnarray}
We emphasize that the normalization factor $1 / \langle\mathrm{DM_{cos}}\rangle$ was omitted in~\citep{2020Natur.581..391M}, leading to a systematic bias in the inferred baryon density parameter. 
It has been found~\citep{2025PhRvD.112h3516Z} that the PDF in Eq.~(\ref{eq:P_IGM}) with $\alpha$, $\beta$ and $\sigma_{\mathrm{d}}$ varying with redshift provides more consistency with the mock data compared to those proposed in~\citep{2020Natur.581..391M,2025arXiv250707090K}.

The value of $\langle\mathrm{DM_{cos}}\rangle$ as a function of redshift 
 depends on the adopted cosmological model. For a spatially flat universe, it is given by~\citep{2003ApJ...598L..79I, 2004MNRAS.348..999I, 2014ApJ...783L..35D}:
\begin{eqnarray}\label{eq:DMIGM}
	\langle\mathrm{DM}_{\mathrm{cos}}\rangle(z)=\frac{3c\ \Omega_\mathrm{b0}h^2 100^2}{8\pi G m_pH_0}\int_{0}^{z}\frac{(1 + z')f_{\mathrm{d}}(z')\chi_{e}(z')}{E(z')}{d}z',
\end{eqnarray}
where $c$, $G$, $m_p$ and $H_0$ are the speed of light, the gravitational constant, the proton mass and the Hubble constant, respectively. The parameter $\Omega_\mathrm{b0}h^2$ with $h\equiv H_0/(100\ \kmsMpc)$ is the present physical baryon density, and
$f_\mathrm{d}$ denotes the baryon fraction in the diffuse ionized state, i.e., baryons not contained in stars or cold neutral gas.
$\chi_e(z)=Y_\mathrm{H} \chi_{e,\mathrm{H}}(z)+\frac{1}{2}Y_\mathrm{He}\chi_{e,\mathrm{He}}(z)$ denotes the ratio of the number of free electrons to baryons. Here, $Y_\mathrm{H}\sim 3/4$ and $Y_\mathrm{He}\sim 1/4$ are the hydrogen and helium mass fractions, respectively, and $\chi_{e,\mathrm{H}}$ and $\chi_{e,\mathrm{He}}$ denote the ionization fractions of hydrogen and helium, respectively. 
In our analysis, we set $\chi_{e,\mathrm{H}}$ and $\chi_{e,\mathrm{He}}$ to unity since both hydrogen and helium are fully ionized at low redshifts ($z<3$)~\citep{2009RvMP...81.1405M, 2011MNRAS.410.1096B}, resulting in $\chi_e(z)=7/8$. 
The dimensionless Hubble parameter $E(z) \equiv H(z)/H_0$ for a flat universe is expressed as:
\begin{eqnarray}
	E^2(z) =\Omega_\mathrm{m0}(1 + z)^{3}+\Omega_\mathrm{r0}(1+z)^4+ \left(1 - \Omega_\mathrm{m0}-\Omega_\mathrm{r0}\right) \exp\left[ \int_{0}^{z} \frac{3(1+w(z'))}{1+z'}{d}z' \right],
\end{eqnarray}
where $\Omega_\mathrm{m0}$ and $\Omega_\mathrm{r0}$ are the present matter  and radiation density parameters, respectively, and  
$w(z)$ denotes  the dark energy equation of state.  When $w(z)=-1$, the standard $\Lambda$CDM model is recovered. If $w(z)=w= \text{const.}$, the $w$CDM model is obtained. If $w(z)=w_0+w_a\ z/(1+z)$ with $w_0$ and $w_a$ being two constants,   this yields the $w_0w_a$CDM model.

The joint log-likelihood function for the full FRB sample, used to constrain the baryon fraction in the extragalactic diffuse ionized gas, $f_\mathrm{d}$,  is then written as:
\begin{eqnarray}\label{eq:likelihood}
	\ln\mathcal{L}_\mathrm{FRB}=\sum_{i=1}^{N_\mathrm{FRB}} \ln P_i(\mathrm{DM}_{\mathrm{obs}}|z),
\end{eqnarray}
where $N_\mathrm{FRB}$ is the total number of FRBs, and $P_i$ is the conditional probability distribution of $\mathrm{DM_{obs}}$ for the $i$-th FRB at the given redshift, given by
\begin{eqnarray}\label{eq:Pi}
	P_i(\mathrm{DM}_{\mathrm{obs}}|z)&=&\int_{0}^{+\infty}\int_{\mathrm{DM_{halo}^{min}}}^{\mathrm{DM_{halo}^{max}}} \int_{0}^{ \mathrm{DM}_{\mathrm{ext}} } P_{\mathrm{ISM}}(\mathrm{DM_{ISM}}) \times  P_{\mathrm{halo}}(\mathrm{DM}_{\mathrm{halo}})\\
	&\quad\times& P_{\mathrm{cos}}(\mathrm{DM}_{\mathrm{ext}}-\mathrm{DM}_{\mathrm{host}}|z) 
	\times P_{\mathrm{host}}(\mathrm{DM}_{\mathrm{host}}) \ {d}\mathrm{DM}_{\mathrm{host}}\ {d}\mathrm{DM}_{\mathrm{halo}} {d}\mathrm{DM_{ISM}}. \nonumber
\end{eqnarray}
Here we define a new variable, $\mathrm{DM}_{\mathrm{ext}}\equiv\mathrm{DM}_{\mathrm{obs}}-\mathrm{DM_{ISM}}-\mathrm{DM_{halo}}$, which sets the upper limit of the $\mathrm{DM}_{\mathrm{host}}$ integral and depends on the observed value $\mathrm{DM}_{\mathrm{obs}}$ of the $i$-th FRB as well as the integration variables $\mathrm{DM_{halo}}$ and $\mathrm{DM_{ISM}}$. When $\mathrm{DM}_{\mathrm{ext}}<0$, the integrand is set to zero to ensure physical validity.
The integration over $\mathrm{DM}_{\mathrm{halo}}$ is performed from 20 to $50~\pccm$ for $P_\mathrm{halo}=\mathcal{N}(35,15^2)$ and from 50 to $80~\pccm$ for $P_\mathrm{halo}=\mathcal{N}(65,15^2)$. 

We use a sample of 124 FRBs, constructed by combining the 115 events compiled by \cite{2025A&A...698A.215G} with 9 additional FRBs. These include FRB 20220222C, FRB 20220224C, FRB 20230125D, FRB 20230613A, FRB 20230907D, and FRB 20231020B from~\citep{2026MNRAS.545f2144P}; FRB 20231128A and FRB 20231204A from~\citep{2025ApJS..280....6C};  and FRB 20241228A from~\citep{2026ApJ...998...97C}. 
We remove events with redshifts below $z=0.02$, since the estimates of the parameters $\sigma_{\mathrm{d}}$, $\alpha$, and $\beta$ in Eq.~(\ref{eq:P_cosmic}) are only valid over the redshift range $0.02$--$3$.
After applying these criteria, 120 FRBs are retained in our analysis.

\section{Other Cosmological Observations}\label{sec:3}

The baryon fraction in the extragalactic diffuse ionized gas, $f_\mathrm{d}$, exhibits degeneracies with other cosmological parameters, particularly the physical baryon density $\Omega_{\mathrm{b0}}h^2$ and the Hubble constant $H_0$ (see Eq.~(\ref{eq:DMIGM})).  To break these degeneracies and obtain tighter constraints, we incorporate additional cosmological observations, namely, SN~Ia, BAO, and CMB data, into our analysis.

\subsection{Type Ia Supernovae}

For SN~Ia data, we adopt the PantheonPlus compilation, which includes 1701 light curves corresponding to 1550 distinct SN~Ia, spanning redshifts from about $0.001$ to $2.26$ \citep{2022ApJ...938..110B,2022ApJ...938..113S}.
This dataset   provides the redshift $z$, corrected distance modulus $\mu_\mathrm{obs}^\mathrm{corr} = m_{B,\mathrm{obs}}^\mathrm{corr} - M_B$, and the associated covariance matrix $\bm{\mathrm C}_\mathrm{SN}$, where $m_{B,\mathrm{obs}}^\mathrm{corr}$ is the corrected apparent magnitude in the B-band, and $M_B$ is the fiducial absolute B-band magnitude.
For the redshift measurements in this dataset, we use the Hubble-diagram redshift $z_\mathrm{HD}$, which accounts for the CMB-frame redshift $z_\mathrm{CMB}$ and corrections from peculiar velocities. To minimize uncertainties associated with peculiar motion, we exclude SN~Ia with $z < 0.01$, following the prescription in~\citep{2022ApJ...938..110B}. This cut leaves 1590 data points, and the corresponding covariance matrix $\bm{\mathrm C}_\mathrm{SN}$ is reduced to a $1590 \times 1590$ submatrix.
The theoretical  distance modulus is given by  
\begin{eqnarray}
	\mu_\mathrm{th}(z)=25+5\log_{10} \left( \frac{D_L(z)}{\mathrm{Mpc}}\right)
\end{eqnarray}
with $D_L(z)=\frac{c}{H_0}(1+z)\int_{0}^{z}\frac{\mathrm{d}z'}{E(z')}$ being  the luminosity distance.  

The joint log-likelihood function for SN~Ia dataset can be obtained through 
\begin{eqnarray}
	\ln\mathcal{L}_\mathrm{SN}=-\frac{1}{2}\chi^2_\mathrm{SN}+\mathrm{const.},
\end{eqnarray}
where $
\chi^2_\mathrm{SN}=\bm{Q}^\dagger {\mathbf{C}}^{-1}_\mathrm{SN} \bm{Q}$, and  $\bm{Q} \equiv (\bm{m}_{B,\mathrm{obs}}^\mathrm{corr}-M_B )-\bm{\mu}_\mathrm{th}(z_\mathrm{HD})$ 
represents the vector of residuals between observed and theoretical distance moduli. The absolute magnitude $M_B$ is treated as a nuisance parameter and is marginalized over in the analysis.

\subsection{Baryon Acoustic Oscillations and Cosmic Microwave Background}\label{sec:BAO_CMB}

For the BAO dataset, we use the latest measurements from the DESI Data Release~2 (DR2)~\citep{2025PhRvD.112h3515A}. Based on various tracer populations, including the Bright Galaxy Sample (BGS), Luminous Red Galaxies (LRG), Emission Line Galaxies (ELG), quasars (QSO), and the Ly$\alpha$ forest, DESI provides 13 measurements in total, covering the distance ratios $D_M/r_d$, $D_H/r_d$, and $D_V/r_d$. The relevant distance quantities are defined as
\begin{eqnarray}
	D_M(z)&=&\frac{c}{H_0}\int_{0}^{z} \frac{\mathrm{d}z'}{E(z')},\\
	D_H(z)&=&\frac{c}{H_0 E(z)},
\end{eqnarray}
and 
\begin{eqnarray}
	D_V(z)=(z D_M(z)^2D_H(z))^{1/3}.
\end{eqnarray}
The comoving sound horizon at the drag epoch $r_d$ is given by
\begin{eqnarray}\label{eq:rs}
r_d = \int_{z_d}^\infty \frac{c_s(z')}{H_0 E(z')} \, dz',
\end{eqnarray}
with $c_s(z)$ denoting the sound speed in the photon-baryon fluid.
The joint log-likelihood function of the BAO data is
\begin{eqnarray}
	\ln\mathcal{L}_\mathrm{BAO}&=&-\frac{1}{2}\chi^2_\mathrm{BAO}+\mathrm{const.}\\
	&=&-\frac{1}{2}\left( \bm{A}_\mathrm{obs}-\bm{A}_\mathrm{th} \right)^\dagger \bm{\mathrm{C}}^{-1}_\mathrm{BAO} \left( \bm{A}_\mathrm{obs}-\bm{A}_\mathrm{th} \right)+\mathrm{const.},
\end{eqnarray}
where $\bm{A}_\mathrm{obs}$ is the vector consisting of observed values of $D_M/r_d$, $D_H/r_d$, and $D_V/r_d$, and $\bm{A}_\mathrm{th}$ is the corresponding vector of theoretical predictions, and $\bm{\mathrm{C}}_\mathrm{BAO}$ is the $13\times13$ covariance matrix of the BAO data.

The CMB data used in our analysis come from the Planck satellite. For the likelihood function of CMB data ($\mathcal{L}_\mathrm{CMB}$), we adopt the Planck 2018 high-$\ell$ likelihood for the temperature (TT), polarization (EE), and cross-correlation (TE) spectra, the low-$\ell$ likelihood for TT and EE spectra~\citep{2020A&A...641A...5P}, as well as the Planck lensing likelihood~\citep{2020A&A...641A...8P}.  The theoretical CMB power spectra and background evolution are computed using the \texttt{CAMB} code~\citep{2000ApJ...538..473L}.

We perform Markov Chain Monte Carlo (MCMC) sampling of the full parameter space using the \texttt{Cobaya} framework~\citep{2021JCAP...05..057T}. The total joint log-likelihood is given by
\begin{eqnarray}
	\ln\mathcal{L}_\mathrm{total} = \ln\mathcal{L}_\mathrm{FRB} + \ln\mathcal{L}_\mathrm{SN} + \ln\mathcal{L}_\mathrm{BAO} + \ln\mathcal{L}_\mathrm{CMB}.
\end{eqnarray}

\section{Estimation of the Baryon Fraction in the extragalactic diffuse ionized gas}\label{sec:4}
\subsection{Baseline Results}
\begin{figure}[htbp]
	\centering
	\includegraphics[width=.32\columnwidth]{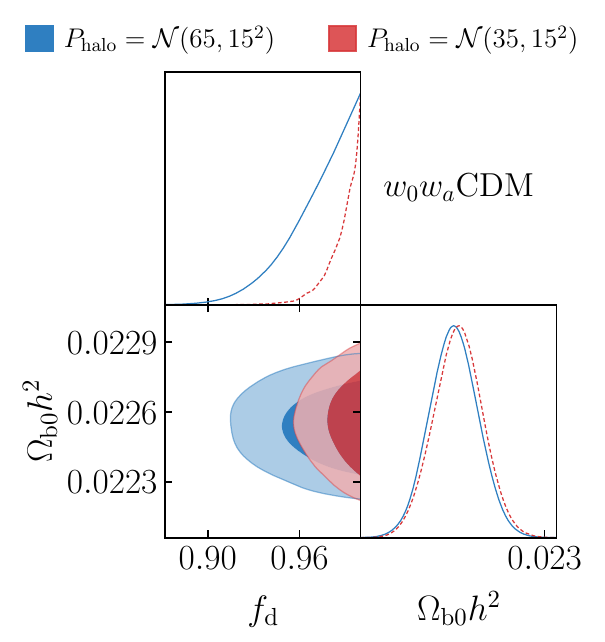}
	\includegraphics[width=.32\columnwidth]{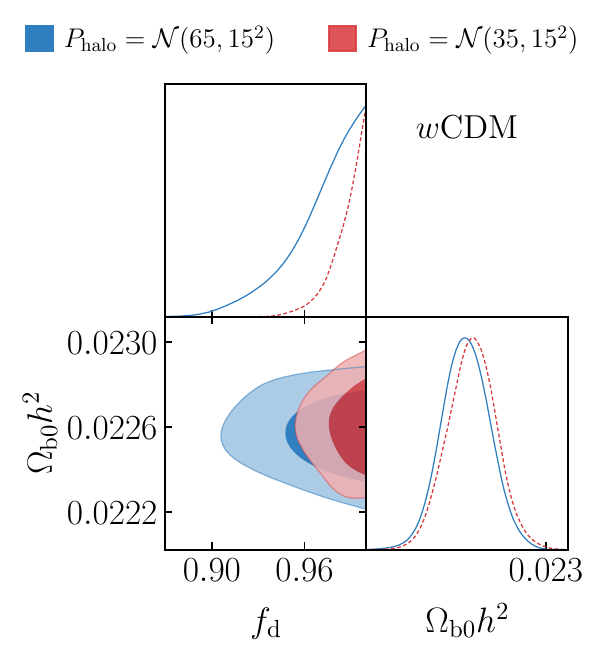}
	\includegraphics[width=.32\columnwidth]{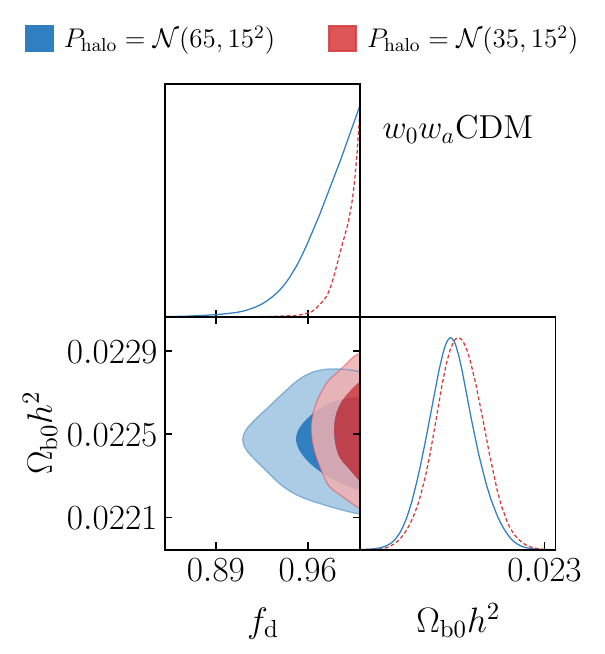}
	\caption{ Constraints on a constant $f_\mathrm{d}$ under different $P_\mathrm{halo}$ models, obtained using the baseline PDF model of $\mathrm{DM}_{\mathrm{cos}}$ (Eq.~(\ref{eq:P_IGM})). Blue solid and red dashed lines correspond to results using the $P_\mathrm{halo}=\mathcal{N}(65,15^2)$ and $P_\mathrm{halo}=\mathcal{N}(35,15^2)$ models, respectively.}
	\label{Fig:1}
\end{figure}

{
We first constrain a constant baryon fraction in the extragalactic diffuse ionized gas, $f_\mathrm{d}$, within the $\Lambda$CDM, $w$CDM, and $w_0w_a$CDM models, adopting the improved PDF of $\mathrm{DM}_{\mathrm{cos}}$ 
in Eq.~(\ref{eq:P_IGM}) as our baseline model. The corresponding marginalized constraints are shown in Fig.~\ref{Fig:1} and summarized in the upper part of Tab.~\ref{tab:1}. 
A notable feature of the results is that the inferred constraints on $f_\mathrm{d}$ are highly consistent across the three cosmological models. 
For a fixed $P_\mathrm{halo}$, the values of  $f_\mathrm{d}$ obtained in the $\Lambda$CDM, $w$CDM, and $w_0w_a$CDM models differ only slightly.  Thus, the $f_{\mathrm{d}}$ constraints are robust against the choice of dark-energy parametrization, even though the cosmological parameters do shift accordingly across the different  models.  In contrast, the assumed distribution of the Milky Way halo contribution, $P_\mathrm{halo}$, has a much stronger impact on the inferred constraints.  For $P_\mathrm{halo}=\mathcal{N}(65,15^2)$, we obtain $f_\mathrm{d}>0.967$, $>0.966$, and $>0.969$ at the 68\% confidence level (CL) in the $\Lambda$CDM, $w$CDM, and $w_0w_a$CDM models, respectively. 
When we instead adopt $P_\mathrm{halo}=\mathcal{N}(35,15^2)$, the corresponding lower bounds become higher, namely $f_\mathrm{d}>0.986$, $>0.984$, and $>0.987$. 
This trend is expected, since a smaller $\mathrm{DM}_{\mathrm{halo}}$ leaves a larger $\mathrm{DM_{ext}}$ budget to be attributed to the extragalactic diffuse ionized gas component, thereby driving the allowed $f_\mathrm{d}$ toward values closer to unity.
These results indicate that the current FRB data favor a high baryon fraction in the extragalactic diffuse ionized gas, and conservatively imply that more than 90\% of baryons reside in this component in all cases considered here.
}

\begin{figure}[htbp]
	\centering
	\includegraphics[width=.32\columnwidth]{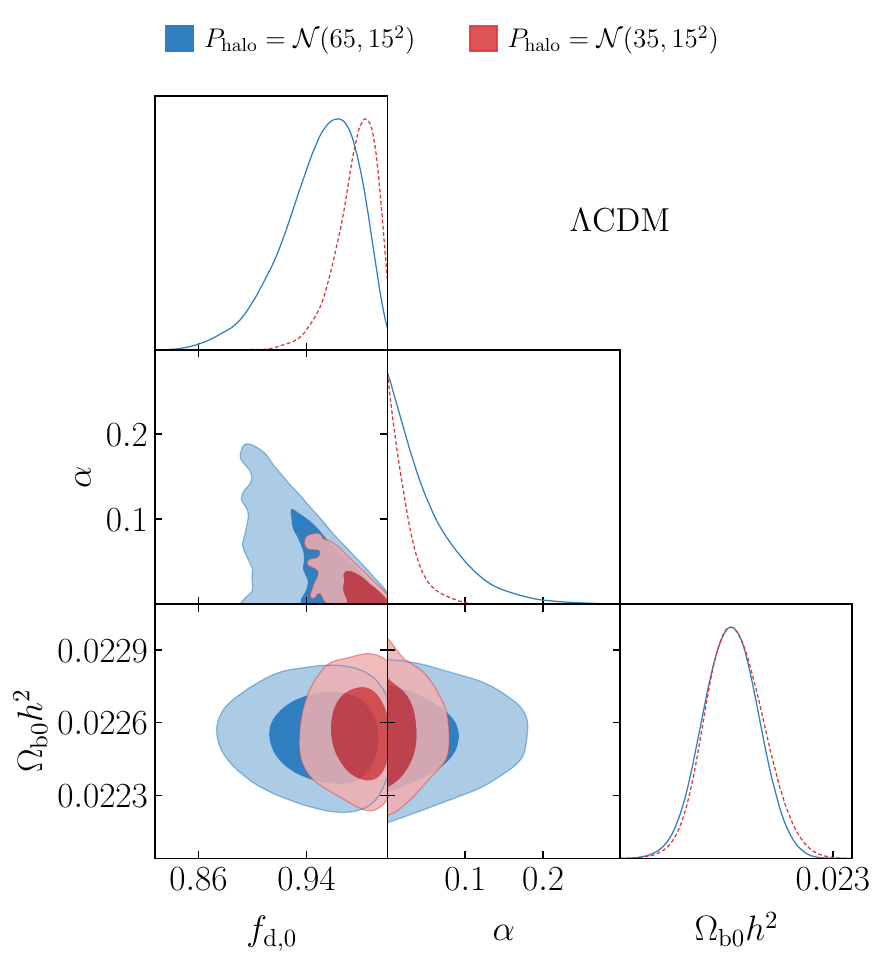}
	\includegraphics[width=.32\columnwidth]{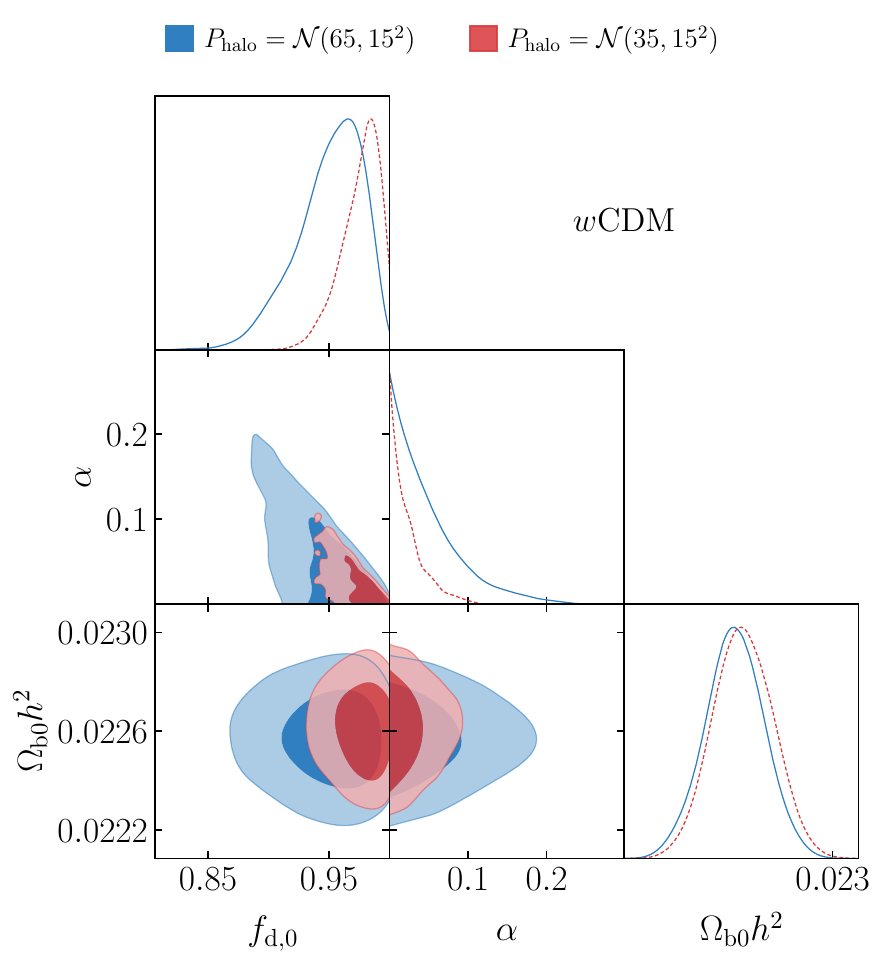}
	\includegraphics[width=.32\columnwidth]{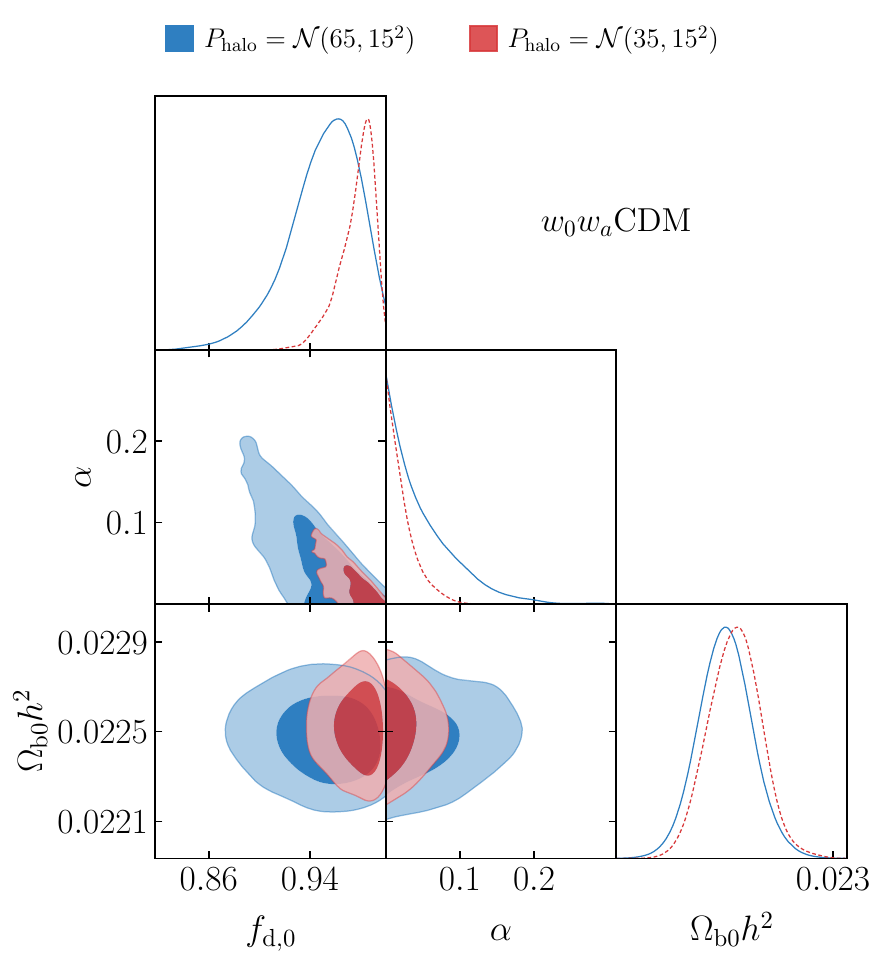}
	\caption{ Similar to Fig.~\ref{Fig:1}, but for the redshift-dependent $f_\mathrm{d}$ with $f_\mathrm{d}(z) = f_{\mathrm{d},0} \left( 1 + \alpha \frac{z}{1+z} \right)$. }
	\label{Fig:2}
\end{figure}

\begin{table}[htbp]
	\centering
	{ 
		\fontsize{8}{10}\selectfont
		\setlength{\tabcolsep}{4pt}  
		\caption{Constraints on $f_\mathrm{d}$ and related cosmological parameters, obtained using the baseline PDF model of $\mathrm{DM}_{\mathrm{cos}}$ (Eq.~(\ref{eq:P_IGM})), for both constant $f_\mathrm{d}$ and redshift-dependent $f_\mathrm{d}$ given in Eq.~(\ref{eq:fdz}) under different cosmological models. The constraints correspond to the 68\% CL; for one-sided constraints, we report lower or upper limits.}
		\begin{tabular}{lcccccccc}
			\hline\hline
			\multicolumn{9}{c}{Constant $f_\mathrm{d}$}\\
			\hline
			Model & $P_\mathrm{halo}$ &$f_\mathrm{d}$ or $f_\mathrm{d,0}$& $\alpha$ & $H_0$ & $\Omega_{\mathrm{m0}}$ & $\Omega_{\mathrm{b0}}h^2$ & $w$ or $w_0$ & $w_a$  \\
			\hline
			$\Lambda$CDM & $\mathcal{N}(65,15^2)$ & $> 0.967$ & - & $68.31\pm 0.28$ & $0.3023\pm 0.0036$ & $0.02254\pm 0.00012$ &- & -\\
			& $\mathcal{N}(35,15^2)$ & $> 0.986$ & - & $68.31\pm 0.29$ & $0.3025\pm 0.0037$ & $0.02256\pm 0.00012$ & - & -\\
			\hline
			$w$CDM				& $\mathcal{N}(65,15^2)$ & $> 0.966$ & - & $67.97\pm 0.58$ & $0.3049\pm 0.0053$ & $0.02257\pm 0.00013$ & $-0.984\pm 0.023$ &-\\
			& $\mathcal{N}(35,15^2)$ & $> 0.984$ & - & $67.79\pm 0.57$ & $0.3062\pm 0.0052$~ & $0.02260\pm 0.00013$ & $-0.976\pm 0.022$ &-\\
			\hline
			$w_0w_a$CDM & $\mathcal{N}(65,15^2)$ & $> 0.969$ & - & $67.58\pm 0.61$ & $0.3110\pm 0.0058$ & $0.02247\pm 0.00014$ & $-0.843\pm 0.055$ & $-0.57^{+0.21}_{-0.19}$ \\
			& $\mathcal{N}(35,15^2)$ & $> 0.987$ & - & $67.44\pm 0.58$ & $0.3121\pm 0.0056$ & $0.02251\pm 0.00014$ & $-0.844\pm 0.053$ & $-0.53\pm 0.20$\\
			\hline
			\hline
			\multicolumn{9}{c}{ Redshift-dependent $f_\mathrm{d}$}\\
			\hline
			$\Lambda$CDM & $\mathcal{N}(65,15^2)$ & $0.949^{+0.035}_{-0.018}$ & $< 0.060$ & $68.32\pm 0.29$ & $0.3023\pm 0.0036$~& $0.02254\pm 0.00012$ &- &-\\
			& $\mathcal{N}(35,15^2)$ & $0.976^{+0.020}_{-0.008}$ & $< 0.025$ & $68.32\pm 0.30$ & $0.3025^{+0.0035}_{-0.0040}$ & $0.02256^{+0.00012}_{-0.00014}$ &- &-\\
			\hline
			$w$CDM				& $\mathcal{N}(65,15^2)$ & $0.950^{+0.035}_{-0.017}$ & $< 0.058$ & $67.95\pm 0.56$ & $0.3050\pm 0.0050$ & $0.02257\pm 0.00013$ & $-0.983\pm 0.023$ &- \\
			& $\mathcal{N}(35,15^2)$ & $0.975^{+0.021}_{-0.008}$ & $< 0.028$ & $67.79\pm 0.58$ & $0.3062\pm 0.0052$ & $0.02261\pm 0.00013$ & $-0.976\pm 0.023$ & -\\
			\hline
			$w_0w_a$CDM  & $\mathcal{N}(65,15^2)$ & $0.950^{+0.034}_{-0.019}$ & $< 0.063$ & $67.52\pm 0.59$ & $0.3115\pm 0.0057$ & $0.02247\pm 0.00013$ & $-0.840\pm 0.056$ & $-0.57\pm 0.21$ \\
			& $\mathcal{N}(35,15^2)$ & $0.977^{+0.017}_{-0.007}$ & $< 0.026$ & $67.39\pm 0.58$ & $0.3125\pm 0.0056$ & $0.02252\pm 0.00013$ & $-0.839\pm 0.053$ & $-0.54^{+0.21}_{-0.18}$ \\
			\hline
			\hline
		\end{tabular}
		\label{tab:1}
	}
\end{table}

To investigate the possible redshift evolution of $f_\mathrm{d}$, we adopt a redshift-dependent parametrization of $f_\mathrm{d}$ ~\citep{2019PhRvD.100j3519W, 2019ApJ...876..146L}
\begin{eqnarray}\label{eq:fdz}
	f_\mathrm{d}(z) = f_{\mathrm{d},0}\left( 1+\alpha \frac{z}{1+z} \right), 
\end{eqnarray}
 where $f_{\mathrm{d},0}$ and $\alpha$ are free parameters. This reduces to the constant case when $\alpha = 0$ and we enforce the physical constraint $f_\mathrm{d}(z) \leq 1$ throughout our analysis.
The constraints on $f_{\mathrm{d},0}$ and $\alpha$ are shown in Fig.~\ref{Fig:2} and summarized in the lower part of Tab.~\ref{tab:1}.

We find that the present value $f_{\mathrm{d},0}$ can be constrained relatively well, whereas the evolution parameter $\alpha$ remains only weakly constrained.
For $P_\mathrm{halo}=\mathcal{N}(65,15^2)$, we obtain $f_{\mathrm{d},0}=0.949^{+0.035}_{-0.018}$, $0.950^{+0.035}_{-0.017}$, and $0.950^{+0.034}_{-0.019}$ at the 68\% CL in the $\Lambda$CDM, $w$CDM, and $w_0w_a$CDM models, respectively. 
For $P_\mathrm{halo}=\mathcal{N}(35,15^2)$, the corresponding constraints become $f_{\mathrm{d},0}=0.976^{+0.020}_{-0.008}$, $0.975^{+0.021}_{-0.008}$, and $0.977^{+0.017}_{-0.007}$. 
As in the constant $f_\mathrm{d}$ case, the inferred values of $f_{\mathrm{d},0}$ are highly consistent across three different cosmological models,  and thus, they  are robust against the choice of dark-energy parametrization.
By contrast, the assumed Milky Way halo contribution still has a noticeable impact, with a smaller halo DM leading to systematically higher values of $f_{\mathrm{d},0}$. 
For the evolution parameter, we obtain only upper bounds: $\alpha<0.060$, $<0.058$, and $<0.063$ for $P_\mathrm{halo}=\mathcal{N}(65,15^2)$, and $\alpha<0.025$, $<0.028$, and $<0.026$ for $P_\mathrm{halo}=\mathcal{N}(35,15^2)$, under the $\Lambda$CDM, $w$CDM, and $w_0w_a$CDM models, respectively. 
These results show that the current FRB data do not provide statistically significant evidence for a redshift evolution of $f_\mathrm{d}$.

To investigate whether the upper limits on $\alpha$ are driven by the specific functional form of $f_\mathrm{d}(z)$ given in Eq.~(\ref{eq:fdz}), we consider an alternative parameterization: $f_\mathrm{d}(z) = f_{\mathrm{d},0}[1 + \alpha z/(1+z)]^{0.5}$, which approaches the physical bound $f_\mathrm{d} \simeq 1$ more gradually. Under the $\Lambda$CDM model, the resulting constraints at 68\% CL are $f_{\mathrm{d},0}=0.949^{+0.036}_{-0.018}$ and $\alpha<0.121$ for $P_\mathrm{halo}=\mathcal{N}(65,15^2)$, and $f_{\mathrm{d},0}=0.975^{+0.021}_{-0.009}$ and $\alpha<0.056$ for $P_\mathrm{halo}=\mathcal{N}(35,15^2)$.  
Compared with the results derived from the parametrization of Eq.~(\ref{eq:fdz}), the $f_{\mathrm{d},0}$ posteriors are nearly unchanged, while the upper limits on $\alpha$ are larger, as expected due to the more gradual evolution form.
However, $\alpha$ remains consistent with zero, and  there is no statistically significant evidence for redshift evolution. It is important to note that  all constraints on   $\alpha$ are limited by the redshift distribution of the sample, which  is concentrated at $z<1$.

The constraints on the other cosmological parameters, including $H_0$, $\Omega_{\mathrm{m0}}$, $\Omega_{\mathrm{b0}}h^2$, $w$ (or $w_0$), and $w_a$, are also summarized in Tab.~\ref{tab:1}. 
We find that these parameters are only weakly affected by the choice of $P_\mathrm{halo}$ and by whether $f_\mathrm{d}$ is modeled as a constant or as a redshift-dependent quantity. 
In particular, the inferred values of $H_0$, $\Omega_{\mathrm{m0}}$, and $\Omega_{\mathrm{b0}}h^2$ remain highly consistent across all cases considered here, indicating that the uncertainty in the Milky Way halo contribution mainly propagates into the determination of $f_\mathrm{d}$ rather than affecting the cosmological parameters.  Similarly, extending the constant $f_\mathrm{d}$ model to the redshift-dependent parameterization does not produce any significant shift in the cosmological constraints.  Figs.~\ref{Fig:1} and \ref{Fig:2}  present respectively the 2-dimensional posterior contours of $\Omega_{\mathrm{b0}}h^2-f_\mathrm{d}$ and  $\Omega_{\mathrm{b0}}h^2-f_\mathrm{d,0}$,  indicating  no significant correlation between them.   Therefore, the degeneracy between $\Omega_{\mathrm{b0}}h^2$ and $f_\mathrm{d}$ in Eq.~(\ref{eq:DMIGM})   is effectively broken by the joint FRB+SN+BAO+CMB analysis. Consequently, the residual uncertainty in $\Omega_{b0}h^2$ contributes negligibly to the uncertainty in the inferred diffuse baryon fraction. For the parameters of dark energy, the best-fit values of $w$ and $w_0$ are slightly greater than $-1$, but the cosmological constant scenario ($w = -1$) remains allowed within $1\sigma$ CL  for $w$CDM and $3\sigma$~CL for $w_0w_a$CDM. For the dynamical dark energy parameter $w_a$, we find $w_a \simeq -0.6 \pm 0.2$, suggesting a preference for dynamical dark energy at more than 2$\sigma$ significance.  These results are in good agreement with those reported by the DESI collaboration~\citep{2025PhRvD.112h3515A}, confirming that the inclusion of FRB data in combination with BAO, CMB, and SN~Ia datasets provides strong constraints on $f_\mathrm{d}$ while leaving other cosmological parameters largely unaffected.

\subsection{Impact of Baryonic Feedback and Selection Effect}
\begin{figure}[htbp]
	\centering
	\includegraphics[width=.32\columnwidth]{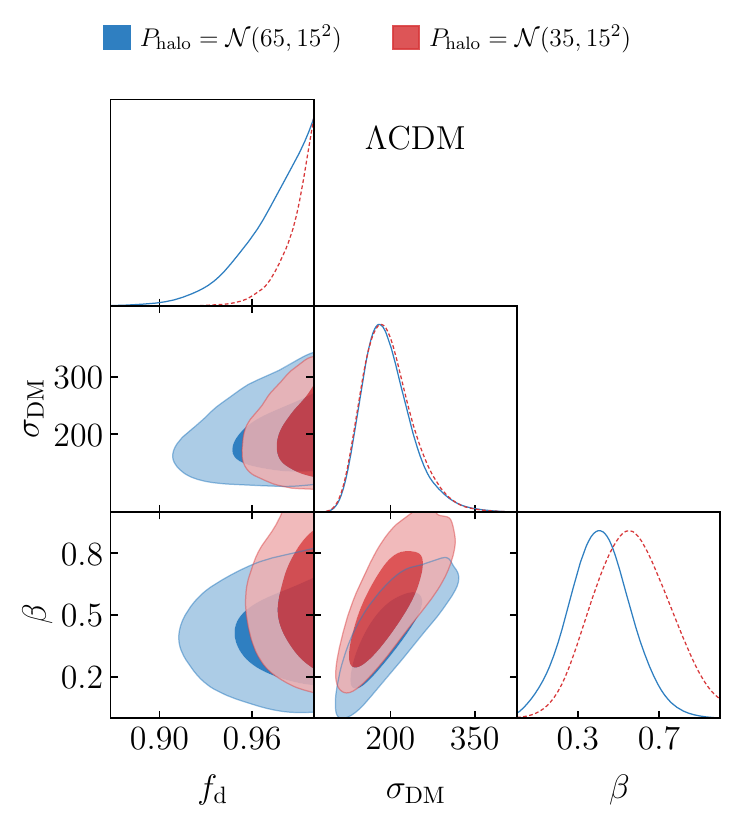}
	\includegraphics[width=.32\columnwidth]{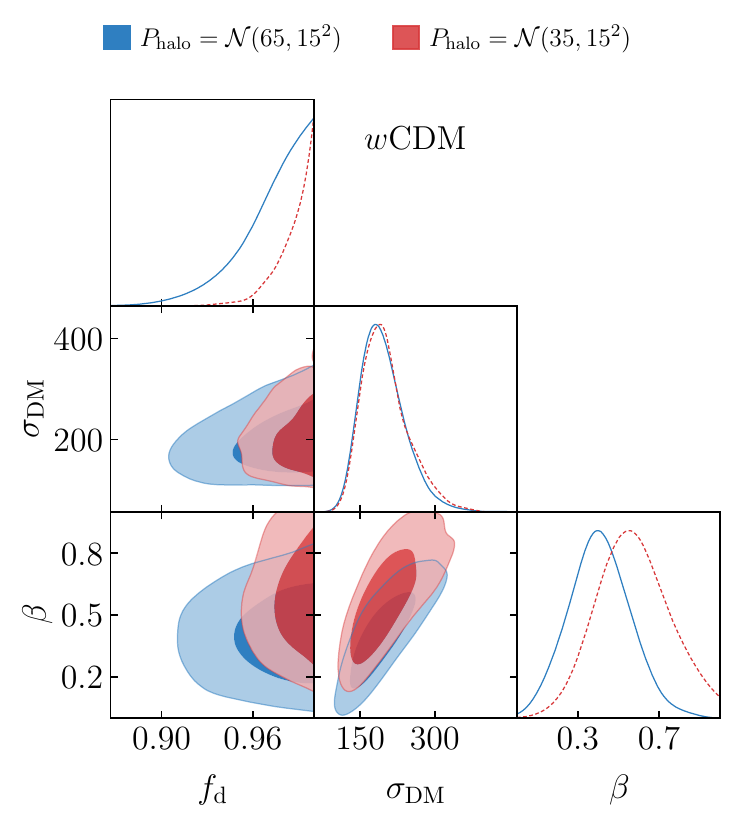}
	\includegraphics[width=.32\columnwidth]{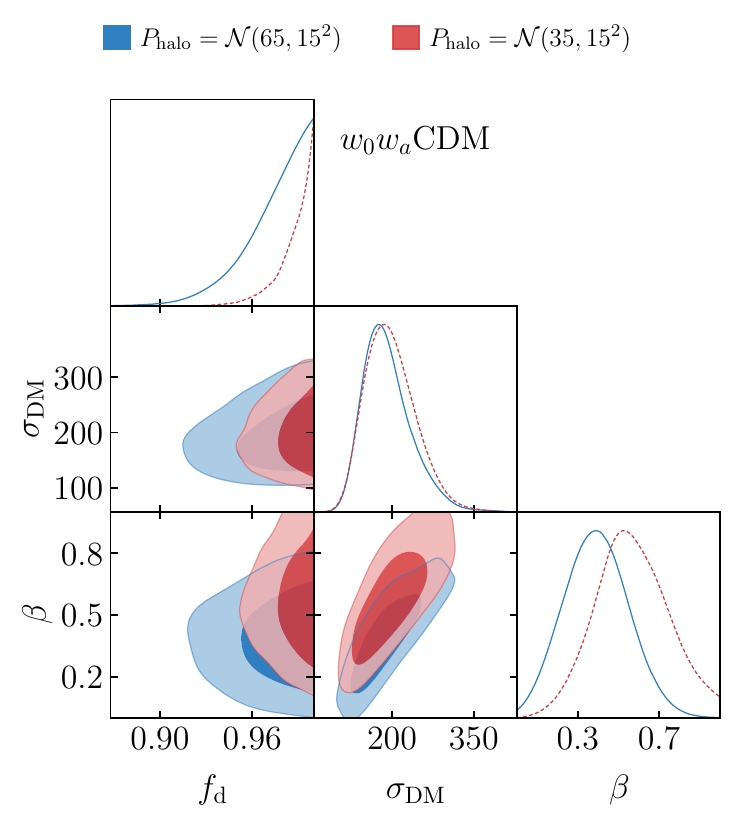}
	\caption{ Constraints on a constant $f_\mathrm{d}$, feedback-related parameters $\sigma_{\rm DM}$ and $\beta$ under different $P_\mathrm{halo}$ models, obtained using the log-normal PDF of $\mathrm{DM}_{\mathrm{cos}}$ (Eq.~(\ref{eq:Pcos_lognormal})). Blue solid and red dashed lines correspond to results using the $P_\mathrm{halo}=\mathcal{N}(65,15^2)$ and $P_\mathrm{halo}=\mathcal{N}(35,15^2)$ models, respectively.}
	\label{Fig:3}
\end{figure}

\begin{figure}[htbp]
	\centering
	\includegraphics[width=.32\columnwidth]{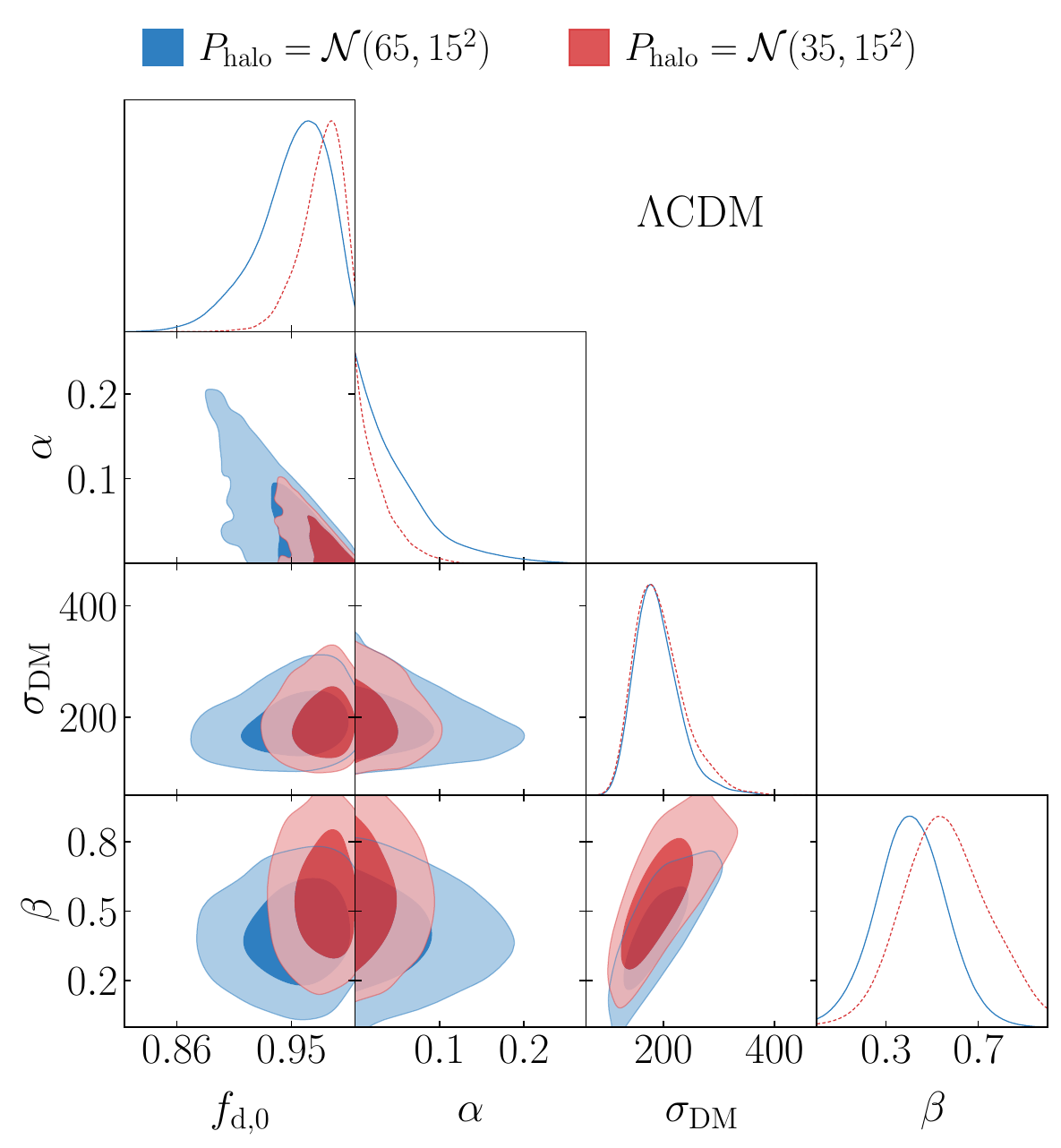}
	\includegraphics[width=.32\columnwidth]{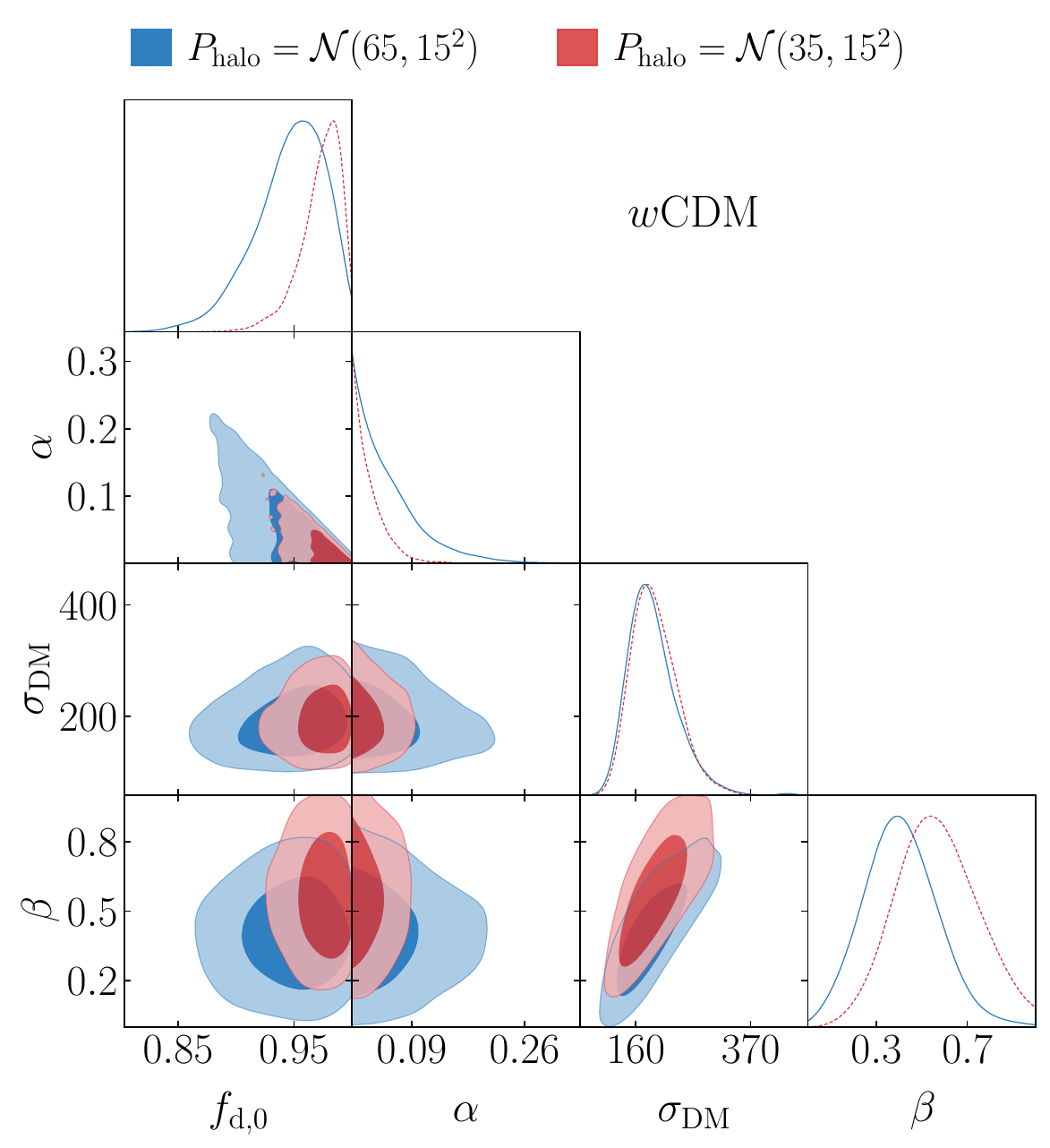}
	\includegraphics[width=.32\columnwidth]{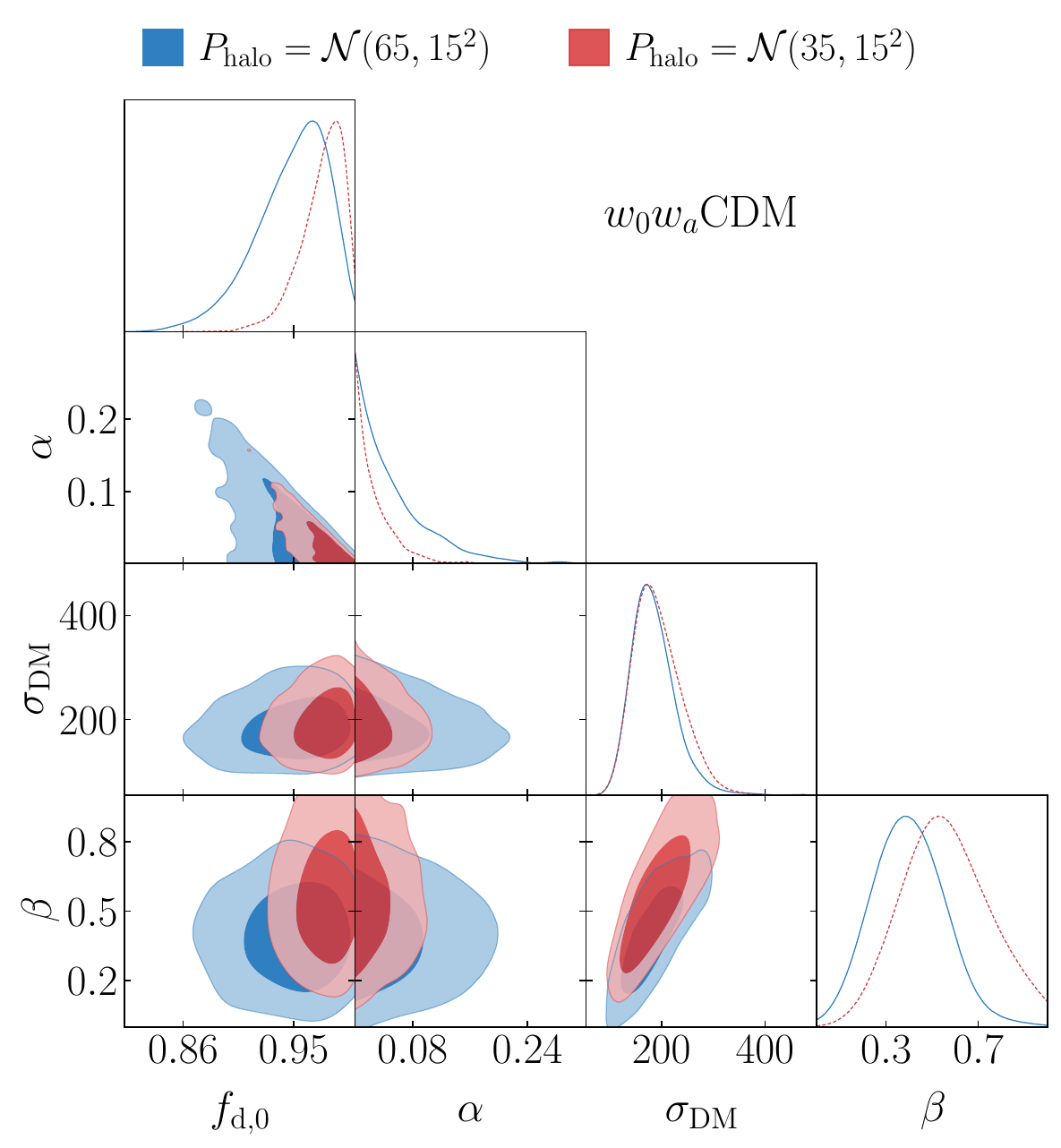}
	\caption{ Similar to Fig.~\ref{Fig:3}, but for the redshift-dependent $f_\mathrm{d}$ with $f_\mathrm{d}(z) = f_{\mathrm{d},0} \left( 1 + \alpha \frac{z}{1+z} \right)$. }
	\label{Fig:4}
\end{figure}

{
We note that the PDF of $\mathrm{DM_{cos}}$ adopted in our analysis (Eq.~(\ref{eq:P_IGM})) is calibrated to specific mock data~\citep{2025arXiv250707090K}, and its shape may be affected by different baryonic feedback strengths. 
We therefore further consider an alternative PDF model for $\mathrm{DM_{cos}}$ to assess the impact of baryonic feedback on the inferred $f_{\rm d}$. 
Using hydrodynamical simulations, \citet{2025ApJ...989...81S} recently found that a log-normal parameterization provides an improved description of $P_{\rm cos}(\mathrm{DM_{cos}}|z)$, and that different feedback prescriptions can lead to different variances of $\mathrm{DM_{cos}}$. Their results suggest that weaker feedback is associated with a broader $\mathrm{DM_{cos}}$ distribution, whereas stronger feedback tends to produce a narrower one. Motivated by this behavior, we also adopt a log-normal model for $\mathrm{DM_{cos}}$~\citep{2025ApJ...989...81S}:
\begin{eqnarray} \label{eq:Pcos_lognormal}
	P_{\rm cos}(\mathrm{DM_{cos}}|z)=
	\frac{1}{\sqrt{2\pi}\,s_\mathrm{cos}(z)\,\mathrm{DM_{cos}}}
	\exp\left[-\frac{\left(\ln \mathrm{DM_{cos}}-\mu_\mathrm{cos}(z)\right)^2}{2s_\mathrm{cos}^2(z)}\right],
\end{eqnarray}
where $s_\mathrm{cos}^2(z) = \ln\left[1+\left(\frac{\sigma_{\rm cos}(z)}{\langle \mathrm{DM_{cos}} \rangle}\right)^2\right]$ and $\mu_\mathrm{cos}(z) = \ln \langle \mathrm{DM_{cos}} \rangle - \frac{1}{2}s_\mathrm{cos}^2(z)$. Here, $\sigma_{\rm cos}$ denotes the standard deviation in linear $\mathrm{DM_{cos}}$ space, which we parameterize as $\sigma_{\rm cos}(z)=\sigma_{\rm DM}\, z^\beta$. This parameterization form is motivated by the hydrodynamical simulation results of \citet{2025ApJ...989...81S}.
Here the feedback-related parameter $\sigma_{\rm DM}$ characterizes the amplitude of $\mathrm{DM_{cos}}$ fluctuation induced by baryonic feedback and $\beta$ describes its redshift evolution.
This PDF is adopted to characterize how the strength of baryonic feedback-related parameters ($\sigma_{\rm DM}$ and $\beta$) affects the inferred $f_{\rm d}$.
We replace the baseline PDF of $\mathrm{DM_{cos}}$ in Eq.~(\ref{eq:Pi}) with the log-normal model given in Eq.~(\ref{eq:Pcos_lognormal}), and constrain both $\sigma_{\rm DM}$ and $\beta$ simultaneously with $f_{\rm d}$ and the other cosmological parameters.
The resulting constraints on $f_{\rm d}$, $\sigma_{\rm DM}$, and $\beta$ are shown in Figs.~\ref{Fig:3} and~\ref{Fig:4}, and summarized in Tab.~\ref{tab:2}. We do not show the cosmological parameter constraints here, since the previous subsection has already indicated that they are governed primarily by the other observation data rather than by the FRB data.

We find that the inferred constraints on $f_\mathrm{d}$ remain stable after taking the feedback-related parameters into account. For the constant $f_\mathrm{d}$ case, all three cosmological models still give only lower bounds, with $f_\mathrm{d}\gtrsim0.96$ for $P_\mathrm{halo}=\mathcal{N}(65,15^2)$ and $f_\mathrm{d}\gtrsim0.98$ for $P_\mathrm{halo}=\mathcal{N}(35,15^2)$ at the 68\% CL. 
For the redshift-dependent parameterization, the constraints are $f_{\mathrm{d},0}=0.951^{+0.035}_{-0.017}$, $0.946^{+0.038}_{-0.020}$, and $0.948^{+0.037}_{-0.019}$ with $\alpha<0.060$, $<0.066$, and $<0.063$ for $P_\mathrm{halo}=\mathcal{N}(65,15^2)$, and $f_{\mathrm{d},0}=0.973^{+0.021}_{-0.009}$, $0.973^{+0.022}_{-0.009}$, and $0.973^{+0.023}_{-0.009}$ with $\alpha<0.032$, $<0.031$, and $<0.032$ for $P_\mathrm{halo}=\mathcal{N}(35,15^2)$, in the $\Lambda$CDM, $w$CDM, and $w_0w_a$CDM models, respectively.
Compared with the baseline results in Tab.~\ref{tab:1}, the constraints on $f_{\rm d}$ in Tab.~\ref{tab:2} change only slightly.
For the feedback-related parameters themselves, we obtain $\sigma_{\rm DM}\sim 200~\pccm$ and $\beta\sim0.4$--$0.6$.
The posterior distributions (Figs.~\ref{Fig:3} and~\ref{Fig:4}) show that the correlations of $\sigma_{\rm DM}$ and $\beta$ with $f_{\rm d}$ are both very weak.
This indicates that the feedback-related parameters $\sigma_{\rm DM}$ and $\beta$ have little influence on the inferred $f_{\rm d}$ for the current dataset.
Therefore, baryonic feedback does not alter our main conclusion that the baryon fraction in the extragalactic diffuse ionized gas is high and shows no significant redshift evolution.

On the other hand, it is worth noting that FRB surveys are incomplete in DM because of selection effects. We therefore also examine the impact of the DM selection function, $S_{\rm DM}$, on the constraints on $f_{\rm d}$. The DM selection function describes the detection probability of an FRB with a given observed DM after marginalizing over other burst properties~\citep{2021ApJS..257...59C}, and it generally differs among telescopes. Since our dataset combines FRBs detected by a variety of instruments, a complete treatment of the selection function for each telescope would be highly complex. However, because our purpose here is only to assess whether $S_{\rm DM}$ has a significant impact on the inferred $f_{\rm d}$, we consider only a simple form of $S_{\rm DM}$ to approximate the selection function of our dataset. In general, FRBs with higher $\mathrm{DM_{obs}}$ are less likely to be detected. We therefore adopt a sigmoid form, $S_{\rm DM}(\mathrm{DM}_{{\rm obs}})=1/(1+\exp(s(\mathrm{DM_{obs}} -\mathrm{DM_{cut}} )))$, where $\mathrm{DM_{cut}}$ and $s$ are set to $1500~\pccm$ and $0.01$, respectively.

When the DM selection function is included, the probability distribution of $\mathrm{DM_{obs}}$ at a given redshift (Eq.~(\ref{eq:Pi})), is rewritten as
 \begin{eqnarray}
	P_{s,i}(\mathrm{DM}_{{\rm obs}}|z) = \frac{P_i(\mathrm{DM}_{{\rm obs}}|z) S_{\rm DM}(\mathrm{DM}_{{\rm obs}})}{ \int_{0}^{\mathrm{DM_{max}}} P_i(\mathrm{DM}_{{\rm obs}}^\prime|z) S_{\rm DM}(\mathrm{DM}_{{\rm obs}}^\prime)d \mathrm{DM}_{{\rm obs}}^\prime },
\end{eqnarray}
where the upper limit of integration is set to $\mathrm{DM_{max}}=3000~\pccm$.
Replacing $P_i$ in Eq.~(\ref{eq:likelihood}) with this expression, and adopting the $w_0w_a$CDM model together with the log-normal form of $P_{\rm cos}$, we constrain the redshift-dependent $f_{\rm d}$ model.
We obtain $f_{\rm d,0}=0.953^{+0.035}_{-0.015}$ and $\alpha < 0.059$ for $P_\mathrm{halo}=\mathcal{N}(65,15^2)$ at the 68\% CL, and $f_{\rm d,0}=0.974^{+0.021}_{-0.008}$ and $\alpha < 0.031$ for $P_\mathrm{halo}=\mathcal{N}(35,15^2)$ at the 68\% CL. 
These results show no significant change relative to the baseline results in the last two rows of Tab.~\ref{tab:1}, indicating that including the DM selection function does not significantly affect our conclusions. This is also consistent with the recent work of \cite{2026ApJ...999..202S}, who showed that, for the current localized FRB sample size, cosmological inference based on a conditional distribution of $\mathrm{DM_{obs}}$ such as that used here (Eq.~(\ref{eq:Pi})), is not strongly affected by selection effects.

}

\begin{table}[htbp]
	\centering
	{ 
		\fontsize{8}{10}\selectfont
		\setlength{\tabcolsep}{4pt}  
		\caption{Constraints on $f_\mathrm{d}$ and the feedback-related parameters $\sigma_{\rm DM}$ and $\beta$, obtained using the log-normal PDF of $\mathrm{DM}_{\mathrm{cos}}$ (Eq.~(\ref{eq:Pcos_lognormal})), for both constant and redshift-dependent $f_\mathrm{d}$ under different cosmological models. The constraints correspond to the 68\% CL; for one-sided constraints, we report lower or upper limits.}
		\begin{tabular}{lccccc}
			\hline\hline
			\multicolumn{6}{c}{Constant $f_\mathrm{d}$}\\
			\hline
			Model & $P_\mathrm{halo}$ &$f_\mathrm{d}$ or $f_\mathrm{d,0}$& $\alpha$ & $\sigma_{\rm DM}$ & $\beta$   \\
			\hline
			$\Lambda$CDM & $\mathcal{N}(65,15^2)$ & $> 0.967$ & - & $195^{+31}_{-51}$ & $0.41\pm 0.15$ \\
			& $\mathcal{N}(35,15^2)$ & $> 0.985$ & - & $196^{+33}_{-52}$ & $0.56\pm 0.18$  \\
			\hline
			$w$CDM				& $\mathcal{N}(65,15^2)$ & $> 0.966$ & - & $197^{+32}_{-53}$ & $0.41\pm 0.15$  \\
			& $\mathcal{N}(35,15^2)$ & $> 0.983$ & - & $201^{+30}_{-57}$ & $0.57^{+0.17}_{-0.19}$ \\
			\hline
			$w_0w_a$CDM & $\mathcal{N}(65,15^2)$ & $> 0.969$ & - & $191^{+30}_{-53}$ & $0.39\pm 0.16$  \\
			& $\mathcal{N}(35,15^2)$ & $> 0.985$ & - & $197^{+36}_{-53}$ & $0.56\pm 0.18$  \\
			\hline
			\hline
			\multicolumn{6}{c}{ Redshift-dependent $f_\mathrm{d}$}\\
			\hline
			$\Lambda$CDM & $\mathcal{N}(65,15^2)$ & $0.951^{+0.035}_{-0.017}$ & $< 0.060$ & $189^{+30}_{-46}$ & $0.41\pm 0.15$ \\
			& $\mathcal{N}(35,15^2)$ & $0.973^{+0.021}_{-0.009}$ & $< 0.032$ & $194^{+31}_{-54}$ & $0.56^{+0.17}_{-0.20}$ \\
			\hline
			$w$CDM				& $\mathcal{N}(65,15^2)$ & $0.946^{+0.038}_{-0.020}$ & $< 0.066$ & $191^{+30}_{-52}$ & $0.41^{+0.15}_{-0.17}$  \\
			& $\mathcal{N}(35,15^2)$ & $0.973^{+0.022}_{-0.009}$ & $< 0.031$ & $194^{+34}_{-48}$ & $0.56\pm 0.18$ \\
			\hline
			$w_0w_a$CDM & $\mathcal{N}(65,15^2)$ & $0.948^{+0.037}_{-0.019}$ & $< 0.063$ & $185^{+31}_{-48}$ & $0.399^{+0.148}_{-0.167}$  \\
			& $\mathcal{N}(35,15^2)$ & $0.973^{+0.023}_{-0.009}$ & $< 0.032$ & $193^{+35}_{-54}$ & $0.555^{+0.180}_{-0.205}$  \\
			\hline
			\hline
		\end{tabular}
		\label{tab:2}
	}
\end{table}

\subsection{Impact of $\mathrm{DM_{host}}$ Distribution}

Tables~\ref{tab:1} and \ref{tab:2} show that, when the host-galaxy DM distribution is fixed using the IllustrisTNG simulation~\cite{2020ApJ...900..170Z}, the dominant systematic uncertainty in the inferred $f_\mathrm{d}$ arises from the assumed Milky Way halo distribution. Since the host-galaxy DM is considerably less well constrained observationally, this treatment may underestimate the associated systematic uncertainty. We therefore investigate how the inferred $f_\mathrm{d}$ changes when the host-galaxy DM distribution is allowed to vary freely.

For each FRB at redshift $z$, the observed-frame $\mathrm{DM_{host}}$ in Eq.~(\ref{eq:DMobs}) is related to its rest-frame counterpart by $\mathrm{DM_{host}}(z) = \mathrm{DM_{host,0}} / (1+z)$.  We assume that the rest-frame $\mathrm{DM}_{\rm host,0}$  follows a log-normal distribution with free parameters $\mu_{\rm host,0}$ and $s_{\rm host,0}$, which are constrained simultaneously with $f_\mathrm{d}$ and the cosmological parameters. Throughout this analysis, we retain the baseline $P_{\rm cos}$ of Eq.~(\ref{eq:P_IGM}) and ignore the selection-function calibration so that any change in the inferred $f_\mathrm{d}$ originates solely from the host-galaxy model. The analysis is performed for both constant and redshift-dependent $f_\mathrm{d}$ models in the $\Lambda$CDM, $w$CDM, and $w_0w_a$CDM frameworks using both halo distributions.
 
As in the previous subsection, the inferred parameters are nearly independent of the adopted cosmological model. 
When comparing these results with those presented   in Tab.~\ref{tab:1},  we find   that   allowing  the host parameters to vary systematically shifts the inferred $f_\mathrm{d}$ to lower values.
In  the  case of constant $f_\mathrm{d}$, the lower bounds decrease by $\sim 0.02$--$0.03$.
For the redshift-dependent case, the central values of $f_{\mathrm{d},0}$ decrease by $\sim 0.03$--$0.06$, accompanied by larger uncertainties compared to the baseline results.  The upper limits on $\alpha$ increase by a factor of $2$ to $5$, while $\alpha$ remains consistent with zero in all cases.

 Interestingly, allowing the host-galaxy parameters to vary substantially reduces the differences in $f_\mathrm{d}$ produced by the two assumed Milky Way halo models. Consequently, the inferred values of $f_\mathrm{d}$, $f_{\mathrm d,0}$, and $\alpha$ become mutually consistent across the adopted halo distributions. In contrast, the inferred host parameters $\mu_{\rm host,0}$ and $s_{\rm host,0}$ differ by more than $1\sigma$ between the two halo models, indicating a partial degeneracy between the halo and host contributions to the observed DM.

To compare our inferred host-galaxy distribution with the IllustrisTNG prediction, we note that 105 of the 120 FRBs in our sample are non-repeating, with a median redshift of $z\simeq0.21$. Our inferred values, $\mu_{\mathrm{host,0}} \sim 4.2$--$4.3$ for $P_\mathrm{halo}=\mathcal{N}(65,15^2)$ and  $\mu_{\mathrm{host,0}} \sim 4.7$--$4.8$ for $\mathcal{N}(35,15^2)$, are systematically larger than the simulation prediction $\mu_{\mathrm{host,0}} \sim 3.9$, suggesting that IllustrisTNG may underestimate the host-galaxy DM contribution for non-repeating FRBs.  A more detailed comparison between the inferred host parameters and simulation predictions is beyond the scope of present work and will be addressed in a future study.

In summary, allowing the host-galaxy DM distribution to vary weakens the constraints on $f_\mathrm{d}$ and increases the uncertainty in the evolution parameter $\alpha$. Overall, the systematic uncertainty associated with the host-galaxy DM distribution is comparable to that arising from the assumed Milky Way halo DM distribution, indicating that both constitute important sources of systematic uncertainty. Nevertheless, our principal conclusions remain unchanged: the lower bound on constant $f_\mathrm{d}$ remains above $0.9$ at the 68\% confidence level, the inferred $\alpha$ remains consistent with zero, and the constraints on $f_\mathrm{d}$ remain robust against the choice of dark-energy parametrization under the current combination of datasets.

\begin{table}[tbp]
\centering
{
\fontsize{8}{10}\selectfont
\setlength{\tabcolsep}{4pt}
\caption{Constraints on $f_\mathrm{d}$ (or $f_\mathrm{d,0}$), $\alpha$, $\mu_{\rm host,0}$, and $s_{\rm host,0}$, for both constant and redshift-dependent $f_\mathrm{d}$ under different cosmological models and $\mathrm{DM_{halo}}$ distributions. The constraints correspond to the 68\% CL; for one-sided constraints, we report lower or upper limits.}
\begin{tabular}{lccccc}
\hline\hline
\multicolumn{6}{c}{Constant $f_\mathrm{d}$}\\
\hline
Model & $P_\mathrm{halo}$ & $f_\mathrm{d}$ & $\alpha$ & $\mu_{\rm host,0}$ & $s_{\rm host,0}$ \\
\hline
$\Lambda$CDM & $\mathcal{N}(65,15^2)$ & $> 0.948$ & - & $4.23^{+0.22}_{-0.19}$ & $1.18^{+0.13}_{-0.17}$ \\
& $\mathcal{N}(35,15^2)$ & $> 0.956$ & - & $4.73\pm 0.13$ & $0.86^{+0.08}_{-0.11}$ \\
\hline
$w$CDM & $\mathcal{N}(65,15^2)$ & $> 0.951$ & - & $4.21^{+0.23}_{-0.19}$ & $1.18^{+0.12}_{-0.17}$ \\
& $\mathcal{N}(35,15^2)$ & $> 0.956$ & - & $4.73^{+0.15}_{-0.11}$ & $0.86^{+0.10}_{-0.11}$ \\
\hline
$w_0w_a$CDM & $\mathcal{N}(65,15^2)$ & $> 0.953$ & - & $4.23^{+0.22}_{-0.18}$ & $1.17^{+0.12}_{-0.17}$ \\
& $\mathcal{N}(35,15^2)$ & $> 0.958$ & - & $4.75\pm 0.13$ & $0.85^{+0.09}_{-0.11}$ \\
\hline
\hline
\multicolumn{6}{c}{Redshift-dependent $f_\mathrm{d}$}\\
\hline
Model & $P_\mathrm{halo}$ & $f_{\mathrm{d},0}$ & $\alpha$ & $\mu_{\rm host,0}$ & $s_{\rm host,0}$ \\
\hline
$\Lambda$CDM & $\mathcal{N}(65,15^2)$ & $0.914^{+0.056}_{-0.033}$ & $< 0.126$ & $4.34^{+0.22}_{-0.19}$ & $1.13^{+0.12}_{-0.16}$ \\
& $\mathcal{N}(35,15^2)$ & $0.916^{+0.058}_{-0.030}$ & $< 0.118$ & $4.80\pm 0.14$ & $0.84^{+0.08}_{-0.10}$ \\
\hline
$w$CDM & $\mathcal{N}(65,15^2)$ & $0.911^{+0.061}_{-0.032}$ & $< 0.130$ & $4.33^{+0.22}_{-0.18}$ & $1.13^{+0.12}_{-0.17}$ \\
& $\mathcal{N}(35,15^2)$ & $0.924^{+0.053}_{-0.028}$ & $< 0.102$ & $4.79\pm 0.13$ & $0.85^{+0.08}_{-0.11}$ \\
\hline
$w_0w_a$CDM & $\mathcal{N}(65,15^2)$ & $0.919^{+0.057}_{-0.027}$ & $< 0.109$ & $4.34\pm 0.21$ & $1.14^{+0.12}_{-0.16}$ \\
& $\mathcal{N}(35,15^2)$ & $0.928^{+0.052}_{-0.026}$ & $< 0.103$ & $4.79\pm 0.13$ & $0.84^{+0.09}_{-0.10}$ \\
\hline
\hline
\end{tabular}
\label{tab:3}
}
\end{table}

\begin{figure}[htbp]
	\centering
	\includegraphics[width=.32\columnwidth]{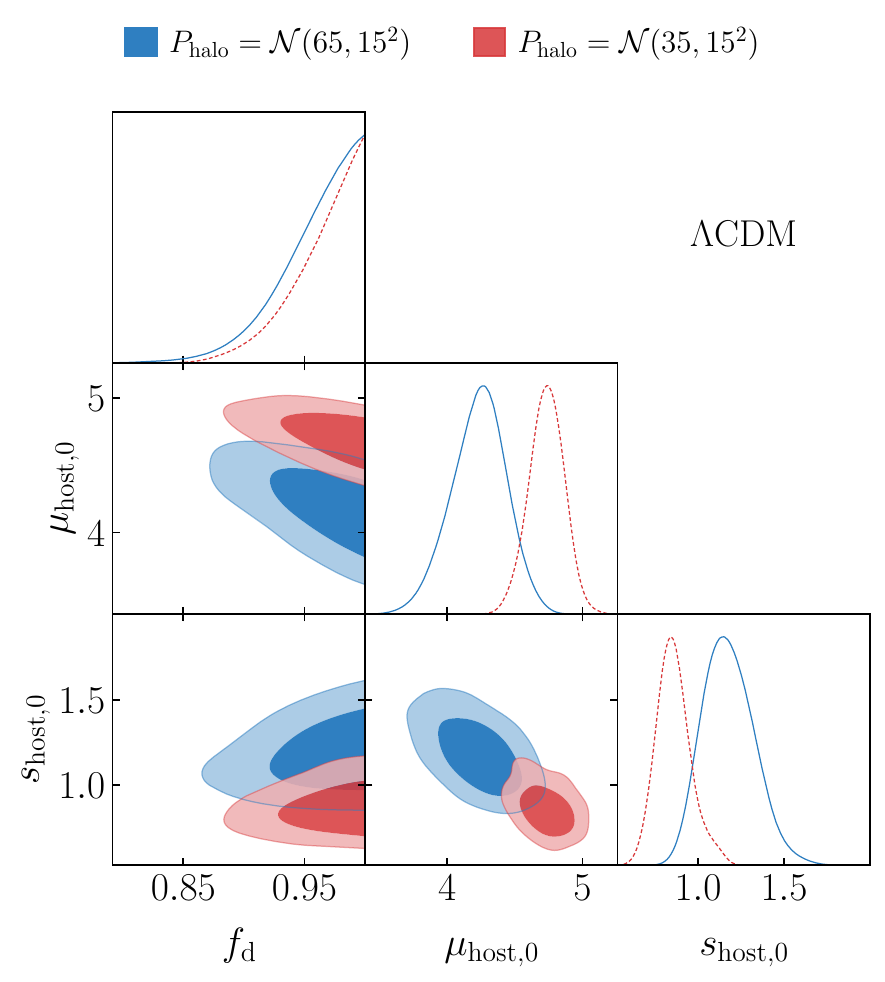}
	\includegraphics[width=.32\columnwidth]{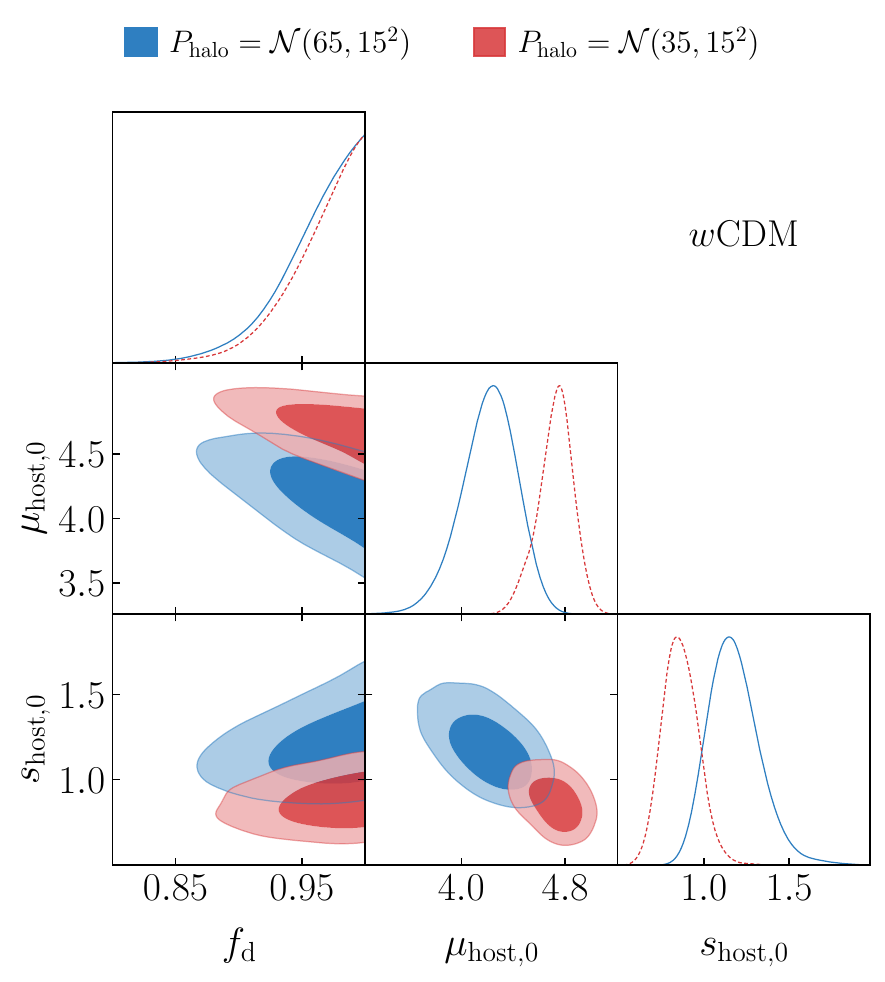}
	\includegraphics[width=.32\columnwidth]{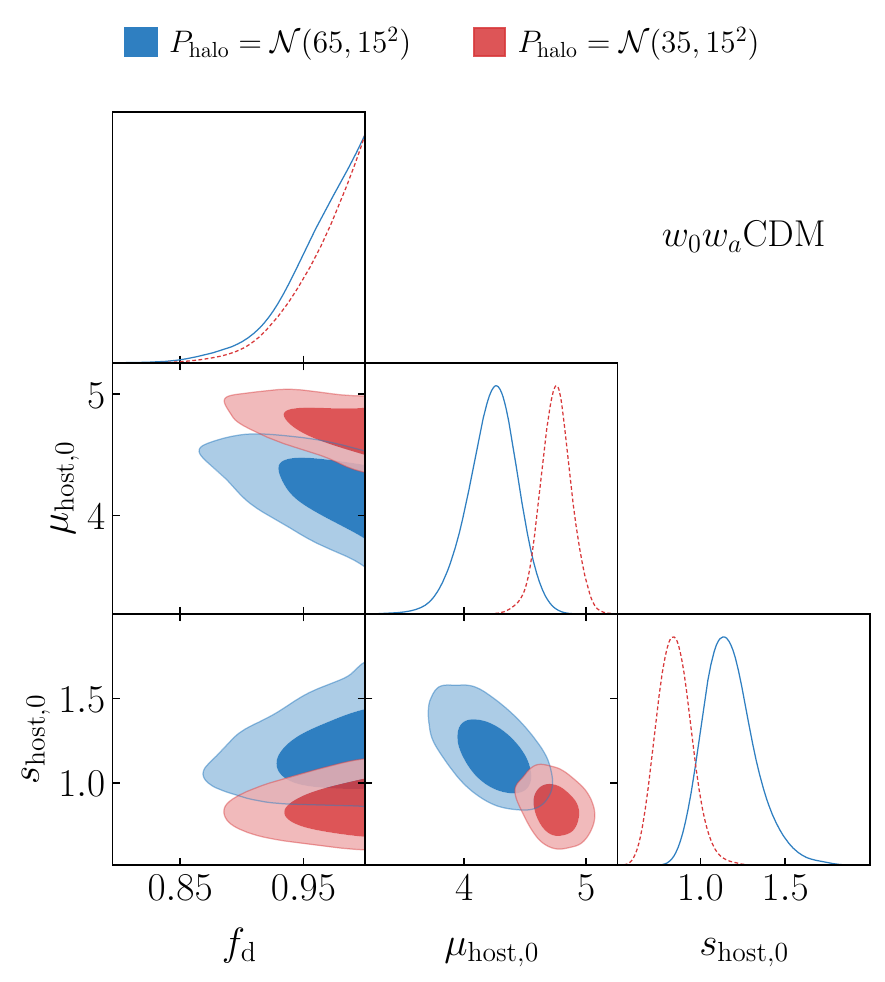}
	\caption{ Constraints on a constant $f_\mathrm{d}$ and the rest-frame host DM distribution parameters ($\mu_{\mathrm{host,0}}$ and $s_{\mathrm{host,0}}$) under different $P_\mathrm{halo}$ models, obtained by treating $\mu_{\mathrm{host,0}}$ and $s_{\mathrm{host,0}}$ as free parameters. Blue solid and red dashed lines correspond to results using $P_\mathrm{halo}=\mathcal{N}(65,15^2)$ and $P_\mathrm{halo}=\mathcal{N}(35,15^2)$, respectively.}
	\label{Fig:5}
\end{figure}

\begin{figure}[htbp]
	\centering
	\includegraphics[width=.32\columnwidth]{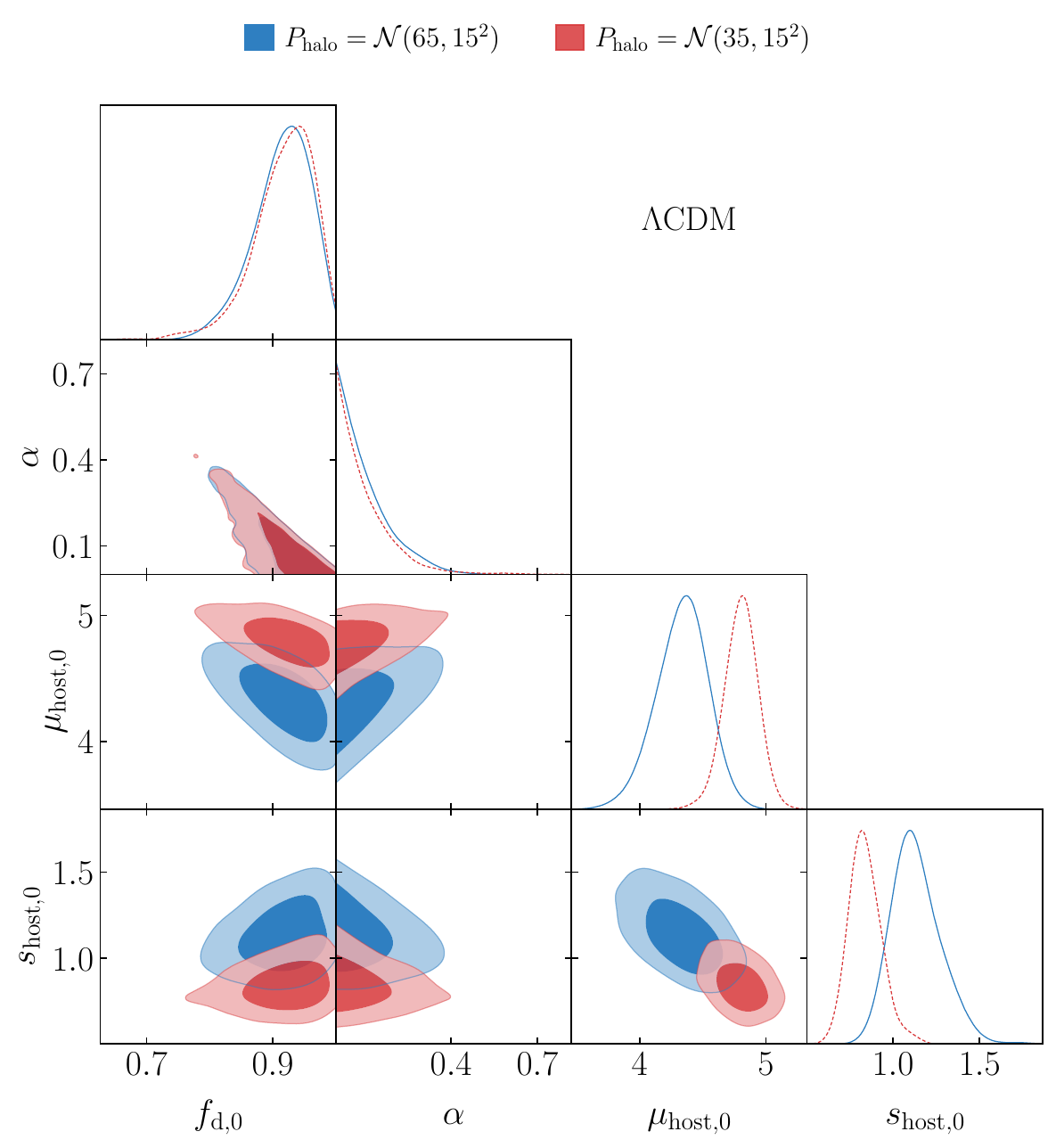}
	\includegraphics[width=.32\columnwidth]{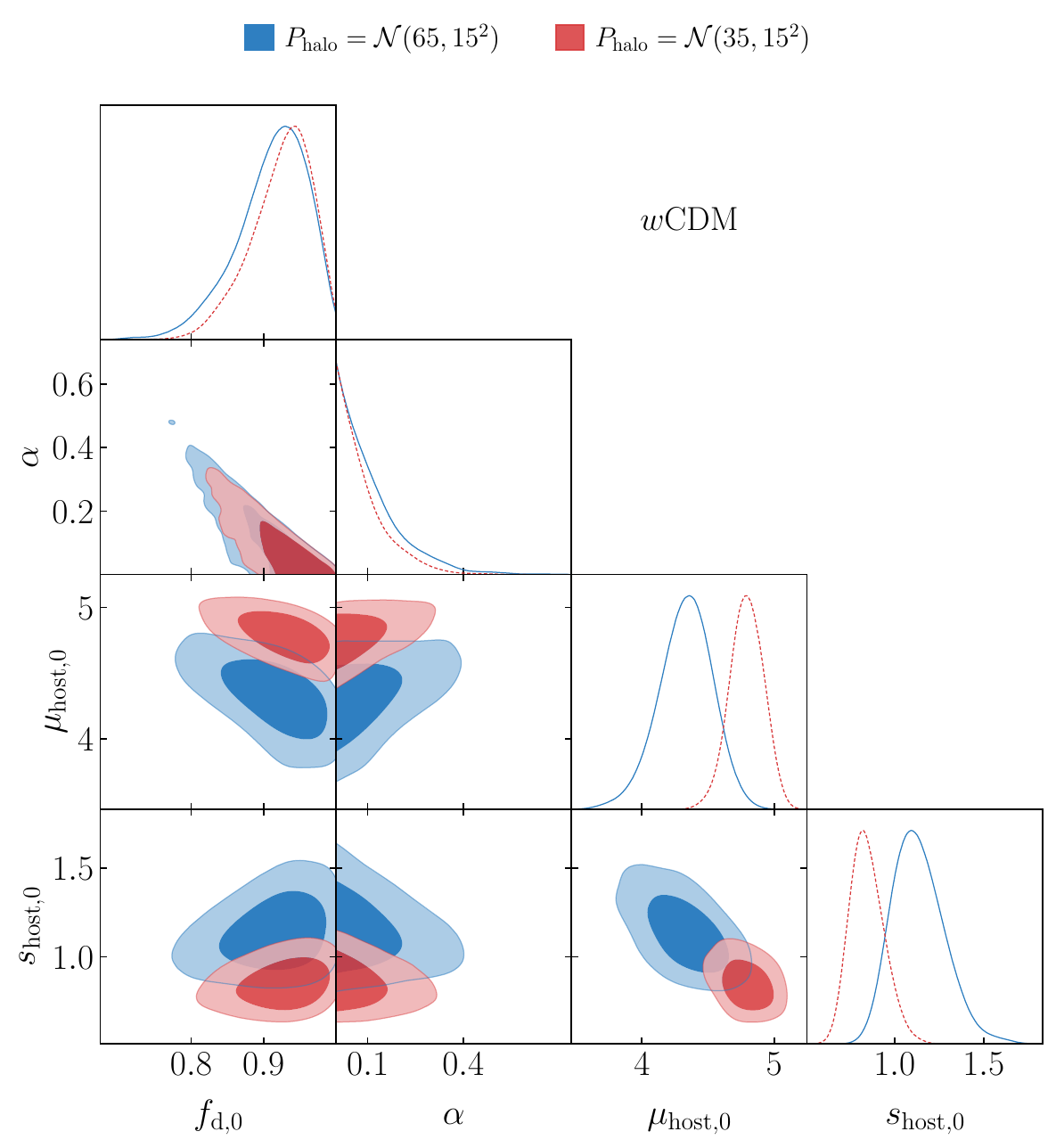}
	\includegraphics[width=.32\columnwidth]{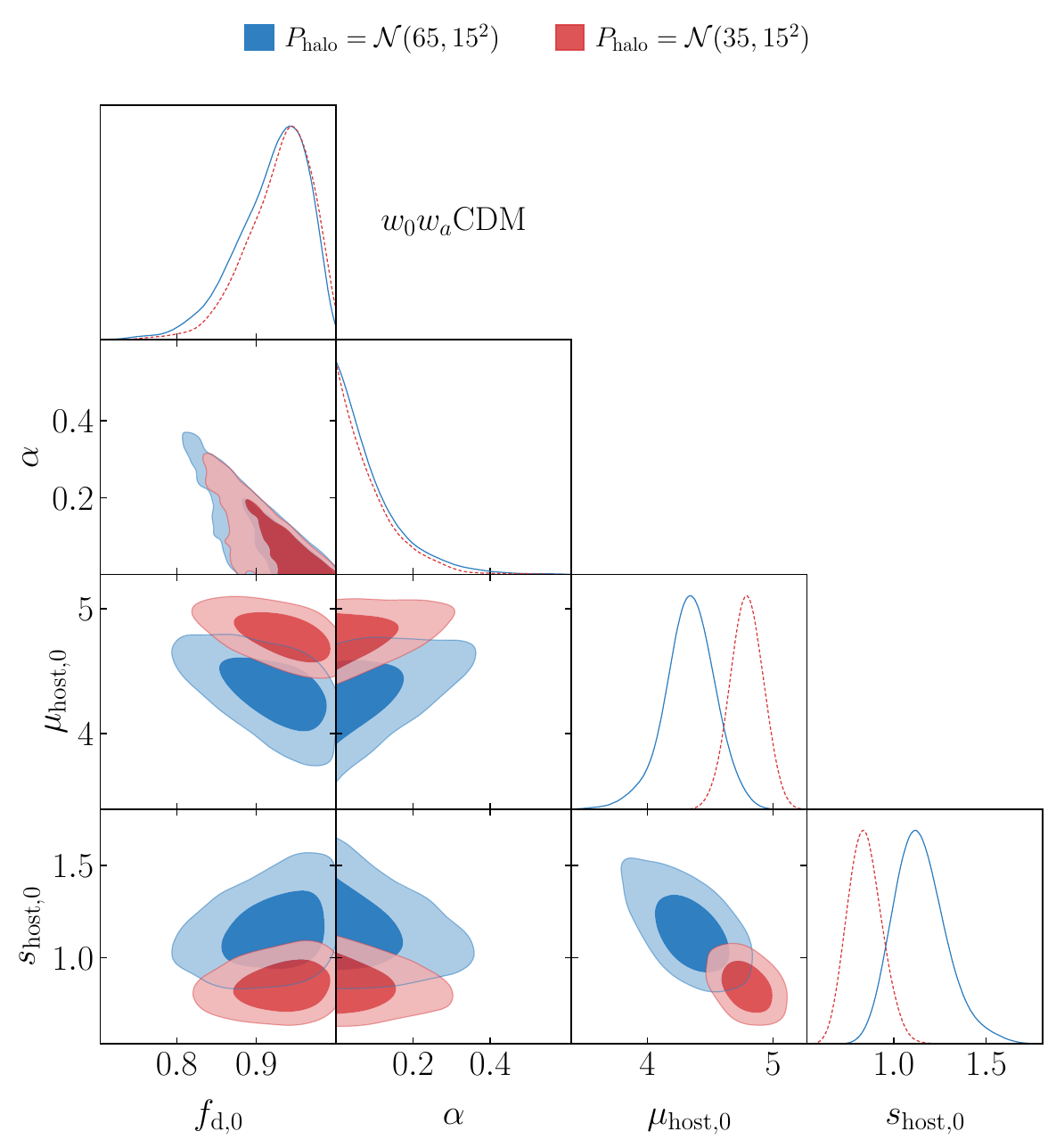}
	\caption{ Similar to Fig.~\ref{Fig:5}, but for the redshift-dependent $f_\mathrm{d}$ with $f_\mathrm{d}(z) = f_{\mathrm{d},0} \left( 1 + \alpha \frac{z}{1+z} \right)$. }
	\label{Fig:6}
\end{figure}

\section{Conclusion}\label{sec:5}

FRBs provide a powerful cosmological probe of the baryon fraction in the extragalactic diffuse ionized gas and thus offer a promising avenue for resolving the long-standing missing baryon problem. In this work, we constrained $f_\mathrm{d}$ using the latest sample of 124 localized FRBs together with the Planck CMB, DESI BAO, and PantheonPlus SN~Ia datasets. Our analysis employed the probability distribution function of $\mathrm{DM}_{\mathrm{cos}}$ proposed in~\citep{2026PhRvD.113d3513Z}, which reproduces mock observations more accurately than previous models. We considered both constant and redshift-dependent parameterizations of $f_\mathrm{d}$ within the $\Lambda$CDM, $w$CDM, and $w_0w_a$CDM cosmologies, while exploring different assumptions for the Milky Way halo and host-galaxy DM distributions.
 
  We find that, the inferred $f_\mathrm{d}$ is robust against the choice of dark-energy parametrization under the current combination of datasets, although the underlying cosmological parameters do shift accordingly. The dominant systematic uncertainties in $f_\mathrm{d}$ instead arise from the assumed Milky Way halo and host-galaxy DM distributions. When the host-galaxy DM distribution is fixed to the IllustrisTNG prediction, the inferred $f_\mathrm{d}$ depends noticeably on the adopted halo model.  Although no statistically significant evidence for redshift evolution of $f_\mathrm{d}$ is found, the current constraints remain limited by the redshift coverage of the available FRB sample. These conclusions remain unchanged after accounting for baryonic feedback and the DM selection function.

 Allowing the host-galaxy DM distribution to vary largely removes the dependence of $f_\mathrm{d}$ on the assumed halo model, albeit with increased statistical uncertainties. At the same time, the inferred host-galaxy parameters shift systematically with the adopted halo model, indicating a partial degeneracy between the halo and host contributions to the observed DM. Consequently, uncertainties in both the Milky Way halo and host-galaxy DM distributions constitute the dominant sources of systematic uncertainty in determining $f_\mathrm{d}$.

Despite these systematic uncertainties, our principal conclusion remains robust. In the constant-$f_\mathrm{d}$ model considered here, more than 90\% of baryons reside in the extragalactic diffuse ionized gas. This result is consistent with the recent independent constraint $f_\mathrm{d}=0.94^{+0.05}_{-0.05}$ reported by~\citet{2025NatAs...9.1226C}, providing further support for the picture that the majority of the missing baryons are contained in the diffuse ionized gas.

\begin{acknowledgements}
We appreciate very much the insightful comments and helpful suggestions by the anonymous referee. Yang Liu and Yuchen Zhang contributed equally to this work. 
This work was supported by the National Natural Science Foundation of China (Grants  No.~12275080, No.~12075084, No.~12321003, No.~12422307, and No.~12373053), the Major basic research project of Hunan Province (Grant No.~2024JC0001), the Innovative Research Group of Hunan Province (Grant No.~2024JJ1006), 
and China Postdoctoral Science Foundation (grant No. 2025M783235).
\end{acknowledgements}

\bibliography{ref_ADS}
\bibliographystyle{apsrev4-1}

\end{document}